\documentclass[epsf,preprint]{aastex}
\shorttitle{The soliton model for pulsar}
\shortauthors{Melikidze et al.}

\begin{document}
\title{The spark-associated soliton model for pulsar radio emission}
\author{George I. Melikidze\altaffilmark{1,2}, Janusz A. Gil\altaffilmark{1}
and Avtandil D. Pataraya\altaffilmark{2}}
\email{gogi@astro.ca.sp.zgora.pl, jag@astro.ca.sp.zgora.pl}
\altaffiltext{1}{J. Kepler Astronomical Center, Pedagogical University, Lubuska
2, 65-265, Zielona G\'ora, Poland}
\altaffiltext{2}{Abastumani Astrophysical Observatory, Al.Kazbegi ave. 2a,
Tbilisi 380060, Georgia}

\begin{abstract}
We propose a new, self-consistent theory of coherent pulsar radio emission
based on the non-stationary sparking model of \citet{rs75}, modified by Gil \&
Sendyk (2000) in the accompanying Paper I. According to these authors, the
polar cap (~with a radius $r_{p}\simeq10^4P^{-0.5}$~cm~) is populated by about
$(r_p/h)^2$ sparks of a characteristic perpendicular dimension $D$
approximately equal to the polar gap height scale $h\thicksim 5\times
10^3P^{3/7}$~cm, separated from each other also by about $h$. Each spark
reappears in approximately the same place on the polar cap for a time scale
much longer than its 10 $\mu$s life-time and delivers to the open magnetosphere
a sequence of $e^{-}e^{+}$ clouds which flow orderly along a flux tube of
dipolar magnetic field lines. The overlapping of particles with different
momenta from consecutive clouds leads to effective two-stream instability,
which triggers electrostatic Langmuir waves at the altitudes of about $50$
stellar radii. This is the only known instability which can develop at the low
altitudes, where the observed pulsar radio emission originates. The
electrostatic oscillations are modulationally unstable and their nonlinear
evolution results in formation of ``bunch-like'' charged solitons. A
characteristic soliton length along magnetic field lines is about $30$~cm, so
they are capable of emitting coherent curvature radiation at radio wavelengths.
A perpendicular cross-section of each soliton at radiation altitudes follows
from a dipolar spread of a plasma cloud with a characteristic dimension near
the star surface of about $D\approx h\approx 50$~meters. The net soliton charge
is about $10^{21}$ fundamental charges, contained within a volume of about
$10^{14}~\rm{cm}^3$. For a typical pulsar, there are about $10^5$ solitons
associated with each of about $25$ sparks operating on the polar cap at any
instant. One soliton moving relativisticaly along dipolar field lines with a
Lorentz factor of the order of 100 generates a power of about $10^{21}$~erg/s
by means of curvature radiation. Then the total power of a typical radio pulsar
can be estimated as being about $10^{27 - 28}$~erg/s. The energy of the soliton
curvature radiation is supported by kinetic energy of secondary
electron-positron plasma created by the primary beam produced by the
accelerating potential drop within the polar gap. A significant fraction of
kinetic energy generated by sparks is radiated away in form of the observed
coherent radio emission.
\end{abstract}

\section{Introduction}

Although more than 30 years have past since the discovery of pulsars, the
mechanism of their radio-emission still remains a mystery. This concerns both
the fundamental problem of coherency, and the specific modulation of pulsar
radiation in the form of micropulses, subpulses and characteristic stable mean
profiles. Ruderman and collaborators in a series of papers
\citep{rs75,cr77a,cr77b,cr80} have attempted to solve for both the mechanism of
coherence of single particle radiation and the organization of emitting
regions. Although their two-stream plasma instability has proven inefficient in
producing observable flux, the latter was partially successful in explaining
details of pulsar radiation modulations. Recently, Gil \& Sendyk (2000;
henceforth Paper I) have modified the spark model of \citet{rs75} and
demonstrated that it explains naturally the single-pulse structure (including a
subpulse drift), the mean profile morphology and polarization. All the observed
characteristics are determined by two basic pulsar parameters: period $P$ and
its derivative $\dot{P}$, along with an observing geometry (inclination and
impact angles). They argued that the pulsar polar cap is populated by a number
of sparks with both characteristic dimension and distance between adjacent
sparks equal approximately to the polar gap height $h \simeq 5\times
10^3~P^{3/7}$ cm, where $P$ is pulsar period in seconds. Therefore, the number
of sparks on the polar cap is determined by the so-called complexity parameter
$a = r_{_{p}}/h$ \citep{gs00}, where $r_{_{p}}\simeq 10^4~P^{-0.5}$ cm is the
canonical polar cap radius. Recently, \citet{dr99} have analyzed with
unprecedented detail the driftbands of subpulses in PSR B0943+10. They were
able to determine both the radiation pattern and the observing geometry
corresponding to a peripheral sightline grazing the stable system of $20$
spark-associated subpulse beams which rotate around the pulsar beam axis. This
finding is the strongest evidence of a non-stationary sparking discharge of
high potential drop in the polar cap acceleration region (polar gap) in radio
pulsars \citep{xqz99,gs00}.

Gil \& Sendyk (Paper I) argued that one spark is anchored to the local pole of
a non-dipolar (presumable sun spot-like) surface magnetic field. This prevents
sparks from fast motion along the planes of field lines towards the pole, which
allows them to reappear in approximately the same places (modulo the slow
${\mathbf E}\times{\mathbf B}$ drift across the planes of field lines) on time
scales much longer than 10 $\mu{\mathrm s}$. If a spark reappears at least
twice at one place on the polar cap,
then a strong Langmuir turbulence should occur due to the two-stream
instability \citep{usov1,usov2,gkm97,am98}. In fact, each spark emits a
sequence of e$^-$e$^+$ plasma clouds flowing orderly along a tube of
spark-associated field lines, which can penetrate each other due to large
spread of momenta. Such a penetration ignites an efficient two-stream like
instability, generating strong Langmuir waves. In this paper we consider a
nonlinear evolution of Langmuir electrostatic oscillations generated by the
two-stream instability within a spark-associated plasma column and show that it
leads to a soliton formation, capable of generating a `bunch-like' coherent
radiation, with the all characteristic features of the observed pulsar radio
emission.

The main and the most decisive problem for coherency of curvature radiation is
the formation and stability of the charged bunches. Despite many attempts there
is no sufficiently well-grounded theory for bunch formation so far. The
curvature radiation of charged particle bunches, themselves produced by a
linear plasma wave, has been proposed as a possible mechanism of pulsar radio
emission \citep[e.g.,][]{rs75, cr77a}. However, as pointed out by
\citet{lmmp86}, bunches produced by a linear electrostatic wave can exists only
over extremely short time scale. As the wave propagates along the magnetic
field line, each fixed spatial point senses the alternating electrostatic field
as time elapses. After a half-period this field changes direction and it begins
to `bunch' particles of the opposite sign. It is thus necessary that the time
scale of the process by which the bunches radiate must be significantly shorter
than half of the plasma oscillation period. At the very least, the condition
$\omega_{_l}<\omega $ must hold (where $\omega_{_l}$ and $\omega $ are the
frequencies of the plasma waves and the waves emitted by bunches,
respectively). On the other hand, for the radiation to be coherent, the linear
characteristic dimension of bunches must be shorter than about half wavelength
of the radiated wave. Since in the linear approximation the dimension of the
bunch is determined by the half wavelength of the plasma waves, another
necessary condition $k_{_l}>k$ should be satisfied (here $k_{_l}$ and $k$ are
the wave vectors of Langmuir waves and curvature radiation, respectively). If
the radiated wave has a `vacuum' spectrum $\omega \approx kc$, then these
conditions are incompatible with each other. Thus, the bunching associated with
high frequency linear plasma oscillations cannot be responsible for the
coherent pulsar radio emission \citep[see also][]{mg99}. In this paper we
propose a new promising model for bunch formation due to slowly-varying
nonlinear plasma processes. Our model is self-consistent and free of
fundamental problems mentioned above.

\citet{am98} have recently studied possible instabilities in the non-stationary
spark-associated magnetospheric plasma. They have found that the two-stream
instability within spark-associated plasma clouds is the only one which can
develop at altitudes below 10\% of the light cylinder, that is at altitudes
about 50 stellar radii $R=10^6$~cm, where pulsar radio emission is expected to
originate \citep[e.g.,][]{kg97,kg98}. As already stated above, Langmuir waves
generated due to these instabilities can not directly produce the observed
pulsar radio emission. In fact, their frequency (~about $100$ GHz~) is much
higher than the observed pulsar radio frequencies \citep[see ][]{am98,mg99}.
Moreover, having an electrostatic nature they can not escape from the plasma.
However, as we will argue in this paper, they can form a charge-separated
solitons. A packet of plasma waves propagating in the relativistic
electron-positron plasma with phase velocities close to (but less than) the
velocity of light is unstable from the modulational standpoint, and its
nonlinear evolution results in the formation of a nonlinear solitary wave
soliton. This process is described by the nonlinear Schr\"{o}dinger equation,
taking into account the nonlinear Landau damping \citep{mp80a,mp80b,mp84}. The
role of the low-frequency perturbations in case of the electron-positron plasma
(that is in the absence of ion sonic waves) is played by the nonlinear beatings
of plasma waves and the nonlinear dumping is determined by the resonant
interaction of the beatings with plasma particles. As we argue in this paper,
in condition prevailing within spark-associated pulsar magnetosphere, Langmuir
soliton can cause an effective charge separation for a period of time
sufficiently long to provide a coherent curvature radiation responsible for the
observed pulsar radio emission.

\section{Coherent curvature radio emission in pulsars}

\subsection{Linear theory}

The properties of the secondary electron-positron plasma created via
\citet{stu71} multiplication process by the primary positrons  depend on the
radius of curvature $\mathcal{R}$ of magnetic field lines above the gap, where
the accelerating electric field is negligible. We assume that $\mathcal{R}$ is
almost equal to the radius of curvature at the
stellar surface $\mathcal{R}_{0}\leqslant 10^{6}$cm \citep{rs75,gs00}. Then,
Sturrock's multiplication factor $\varkappa \geqslant 10^{4}$ and the Lorentz
factor of the secondary $e^{-}e^{+}$ plasma $\gamma_{_p}\leqslant 100$. In
fact, let us consider a canonical maximum potential drop within the gap
\citep{stu71,rs75}
\begin{equation}
V\thicksim 1.7\times 10^{12} P^{0.36} \dot{P}_{_{-15}}^{0.5} \qquad \rm{Volts,}
\label{Eq1}
\end{equation}
which holds for $\mathcal{R}_{6}=\mathcal{R}/10^6\approx 1$, where
$\dot{P}_{_{-15}}$ is period derivative in $10^{-15}\rm{s~s^{-1}}$. Then, the
corresponding Lorentz factor of the primary particles should be
$\gamma_{_b}\approx V\left( e/m_{e}c^{2}\right)\thicksim 3\times 10^{6}$ and
for average Lorentz factor of spark-generated plasma particles we have
$\gamma_{_p}\approx\gamma_{_b} /(2\varkappa)\sim 10^2$. The secondary plasma
with a number density $n_{p}\simeq \varkappa n_{_{GJ}}$ is penetrated by the
beam of primary particles with a Goldreich-Julian corotational number density
\citep{gj69}
\begin{equation}
n_{_{GJ}}\thicksim 5.\,6\times 10^{5}~\left( \frac{\dot{P}_{_{-15}}}{P}
\right)^{0.5}R_{_{50}}^{-3}\qquad \rm{cm^{-3}},
\label{nGJ}
\end{equation}
where $R_{_{50}}=r/(50R)$ is the radial distance in units of $50$ stellar radii
$R=10^6$~cm.

It is now well known that the interaction proposed by many authors
\citep[e.g.,][]{lmmp86,mach91,kmm91,kmm92,lbm99} between the primary beam and
magnetospheric plasma is too weak at low altitudes (say $r \thicksim 50R$),
since the instability development requires that drift velocity of the primary
particles becomes sufficiently high \citep{kmm92}. As we already mentioned,
there is only one instability which can produce a strong initial turbulence at
low altitudes. This instability is triggered by the non-stationary process of
plasma creation associated with sparking discharge of the acceleration region
above the polar cap. Before we start exploring the nonlinear effects, let us
briefly summarize the results of the linear approach performed by \citet{am98}.
The plasma frequency and the frequency of excited plasma waves are respectively
\begin{equation}
\omega_{_p}\thicksim 4.\,2\times 10^{9}~R_{_{50}}^{-1.5}
\varkappa_{4}^{0.5}\left( \frac{\dot{P}_{_{-15}}}{P}\right) ^{0.25}\qquad
\rm{rad~s^{-1}}
\label{wp}
\end{equation}
and
\begin{equation}
\omega_{_l}\thicksim 2~\delta_{\omega}~\gamma_{_p}~\omega_{p}\thicksim
4.2\times 10^{11}~R_{_{50}}^{-1.5}\varkappa_{4}^{0.5}\gamma_{_2}^{0.5}\left(
\frac{\dot{P}_{_{-15}}}{P}\right)^{0.25}\qquad \rm{rad~s^{-1}}.
\label{wl}
\end{equation}
In these expressions, $\delta_{\omega }$ is the parameter which has been
calculated in \citet{am98} and estimated as $\thicksim 0.5$,
$\varkappa_{4}=\varkappa/10^4$, where $\varkappa\sim\gamma_{_b}/\gamma_{p}$,
and $\gamma_{_2}=\gamma_{_p}/100$. Here $R_{_{50}}$ is an altitude of
instability region, and for typical pulsars $R_{_{50}}\thicksim 1$. Also
$\varkappa_{4}$ and $\gamma_{_2}$ are regarded as being close to unity
\citep[e.g.,][]{rs75}. Thus for typical values of pulsar parameters
$\omega_{_p}\thicksim 10^{10}~\rm{rad~s^{-1}}$ and $\omega_{_l}\thicksim
10^{12}~\rm{rad s^{-1}}$.

The Langmuir waves with frequency $\omega_{_l}$ determined by equation
(\ref{wl}) are generated by the following simple mechanism
\citep{usov1,usov2,gkm97,am98}. The repeatable sparking creates a succession of
plasma clouds moving along a tube of magnetic field lines, each cloud
containing particles with a large spread of momenta. Overlapping of particles
with different energies (detemined by $\gamma_{_T}$, see Appendix \ref{Defs}
and Fig.\ref{fig1}) from adjacent clouds ignites strong Langmuir oscillations,
which may lead eventually to the generation of coherent pulsar radio emission.
Interestingly, this instability is the only one which, according to our
knowledge, develops at altitudes of the order of a few percent of the light
cylinder radius, where the pulsar radio emission is expected to originate
\citep{cor78,cor92,kg97,kg98}. The altitude $R_{50}$ at which the two-stream
instability can develop depends on the average Lorentz factor of plasma
$\gamma_{_p}$. This has been estimated by \citet[their Fig. 6]{am98}; see also
a kinematic estimate in equation (\ref{R50}) below. Specifically, if
$\gamma_{_2}=0.5$ then $R_{_{50}}\simeq0.1 - 0.5;$ if $\gamma_{_2}=0.75$ then
$R_{_{50}}\simeq0.3 - 1.1;$ if $\gamma_{_2}=1$ then $R_{_{50}}\simeq0.5 - 2$.

The two adjacent secondary plasma clouds corresponding to the two consecutive
sparks are separated by about $\Delta t\thicksim h/c$ (typically $10^{-7}$ s),
where $h\simeq 5\times 10^{3}\mathcal{R}_{6}^{2/7}B_{12}^{-4/7}P^{3/7}$ cm is
the polar gap height, $\mathcal{R}_{6}=\mathcal{R}/R$ and
$B_{12}=B_{0}/10^{12}$ \citep{rs75}. Let us estimate the time $\Delta T$ after
which particles with different Lorentz factors will overcome each other. The
corresponding velocity difference is determined by the average Lorentz factor
as $\Delta v\simeq c/(2\gamma_{_p}^{2})$. It is easy to show that $\Delta
T\thicksim h/\Delta v\thicksim 2\gamma_{_p}^{2}~h/c$. The distance covered
during this time $r_{_{in}}\thicksim c\Delta T\thicksim
2\gamma_{_p}^{2}h\thicksim
10^{8}\gamma_{_2}^{2}\mathcal{R}_{6}^{2/7}B_{12}^{-4/7}P^{3/7}\gg R$, and thus
one can write the expression
\begin{equation}
R_{50}\thicksim \gamma_{_2}^{2}\mathcal{R}_{6}^{2/7}B_{12}^{-4/7}P^{3/7}.
\label{R50}
\end{equation}
This kinematic estimate of the altitude of the instability region agrees
roughly with estimates of the altitude of radio emission region $r/R=50\cdot
R_{50}\thicksim 50\cdot P^{0.33\pm 0.05}$ given by Kijak \& Gil (1997, 1998).

The linear growth rate $\Gamma_{l}$, which should satisfy the condition
$\Gamma_{l}\gg c/\Delta r$, where $\Delta r\sim 50\cdot R_{50}\cdot 10^6$~cm is
characteristic longitudinal dimension of the instability region, can be written
as
\begin{equation}
\Gamma_{l}\thicksim 1.1\times 10^{6}~\gamma_{_2}^{-1.5}~R_{_{50}}^{-1.5}\left(
\frac{\dot{P}_{_{-15}}}{P}\right)^{0.25} ,
\label{Gl}
\end{equation}
and the above condition for the instability development in the resulting plasma
cloud is
\begin{equation}
\gamma_{_2}^{-1.5}~R_{_{50}}^{-1.5}\left(
\frac{\dot{P}_{_{-15}}}{P}\right)^{0.25}\gg 0.1,
\label{Cl}
\end{equation}
\citep{am98}. It is obvious that for typical values of magnetospheric plasma
parameters ($\gamma_{_{2}}\thicksim R_{_{50}}\thicksim
\dot{P}_{_{-15}}\thicksim P \thicksim 1$) the growth rate of instability is
high enough to provide a strong Langmuir turbulence, that is the condition
(\ref{Cl}) is well satisfied. In the following we will explore a nonlinear
evolution of this turbulence and argue that it results in formation of a
`bunch-like' charged soliton, capable of emitting coherent curvature radiation
at radio wavelengths.

\subsection{The nonlinear theory}

It is obvious that the excitation of longitudinal electrostatic waves still
does not explain the observed radio emission of pulsars. We need some mechanism
which leads to the generation of low frequency waves capable of escaping from
the pulsar magnetosphere in the form of coherent radio emission. \citet{karp75}
were the first to propose Langmuir solitons as a possible bunching mechanism.
The net charge in their soliton was due to mass difference between protons and
electrons. \citet{mp80b,mp84} have studied the same problem in the more
realistic case (for pulsars)of an electron-positron plasma and proposed two
possible reasons for charge separation: (i) small admixture of ions, (ii)
difference in the distribution functions of electrons and positrons. Such a
difference occurs naturally in the case of the pulsar magnetospheric plasma,
which will be discussed later in this paper. Following \citet{mp84},
\citet{ass93} discussed the possibility of curvature radiation of Langmuir
soliton, but without a  consideration of non-stationary character of
electron-positron plasma. The basic soliton parameters like dimensions, volume
and net charge are markedly different in \citet{ass93} as compared with this
paper. Moreover, \citet{ass93} calculates the power emitted by soliton using a
single-particle curvature radiation scheme, which is inapplicable as
demonstrated in this paper (see eq.[\ref{Eq12}]).

A packet of plasma waves propagating through a relativistic electron-positron
plasma with phase velocities close to the velocity of light is unstable from
the modulation standpoint, and its nonlinear evolution results in the formation
of a soliton. This process is described by a nonlinear Schr\"{o}dinger equation
with nonlinear Landau damping \citep{mp80b,mp84}. In the case of
electron-positron plasma  the role of low-frequency perturbations is played by
the nonlinear beatings of plasma waves, and the resonant interaction of
beatings with particles determine the nonlinear damping.

In order to avoid confusion in the following discussion, we outline below the
physics of the modulational instability occurring in the pulsar secondary pair
plasma associated with sparking discharge of the polar gap. As already
mentioned, the two-stream instability triggers linear plasma waves in
electron-positron clouds created by succesive sparks. Obviously, there is a
small spread $\Delta\omega$ of frequencies around the characteristic frequency
$\omega_{_l}$ (eq.[\ref{wl}]) of excited plasma waves. Since
$\omega_{_l}\gg\Delta\omega$, the amplitude of linear wave packet, containing
waves with different frequencies near $\omega_{_l}$, will be modulated by low
frequency beatings. The characteristic phase velocity of beatings $\Delta\omega
/\Delta k$ is approximately equal to the group velocity of linear plasma waves
$v_{_g}=d\omega/dk$. Therefore, a resonant interactions of plasma particles
with low frequency beatings (see the resonant factor $(v-v_{_g})^{-1}$ in
eqs.[\ref{Def_W}] - [\ref{Def_U}]) will result in the modulational instability.
Those low frequency beatings which are in resonance with plasma particles will
affect the amplitude of linear waves in the same way as the low frequency
ion-sonic waves which affect the amplitude of Langmuir waves in the well-known
laboratory electron-ion plasma \citep{zs72}.

In places where the amplitude of linear waves increase/decrease the plasma
density $n_{_p}$ decreases/in\-creases and, as a consequence, the
characteristic frequency $\omega_{_l}\propto(n_{_p})^{1/2}$ decreases/increases
as well. On the other hand, the phase velocity of linear waves
$v_{_f}=\omega_{_l}/k$ increases/decreases with increasing/decreasing frequency
or with decreasing/increasing waves amplitude. Therefore, a spread of phase
velocities along wave packet exists. Let us consider a point where the wave
amplitude is a minimum and thus the phase velocity is a maximum at a given
instant. The wave numbers change differently in different directions in the
vicinity of this point. In fact, in the direction of waves propagation
wavelengths $\lambda$ will be shortened and in the opposite direction
wavelengths will be lengthened. This means thats starting from the point of
minimum amplitude, the wave numbers $k=2\pi/\lambda$ increase in the direction
of wave propagation and decrease in the opposite direction.

The behaviour of the modulationally unstable wave packet is determined by the
so-called Lighthill condition \citep{lghill}, which examines the sign of the
product of coefficient $G$ (describes dispersion; eq. [\ref{Def_G}]) and $q$
(describes grow of nonlinearity; eq. [\ref{Def_q}]). If $Gq>0$ then  at places
where the linear waves amplitude is  near maximum, it will grow even larger.
This should lead to self-condensation of the wave packet or soliton formation.
Inspection of Figure \ref{fig1} shows that in pulsar magnetospheres both $G$
and $q$ are positive in wide range of parameters (see eq.[\ref{G_s_q}] and
Appendix \ref{Defs} for definitions) and thus we can adopt that $Gq>0$ in radio
pulsars.

Let us check what effect the positive value of $G\propto dv_{_g}/dk$
(eq.[\ref{Def_G}]) has on the system of linear plasma waves in the vicinity of
the point with minimum wave amplitude. Recall that wave numbers increase on one
side (in the direction of propagation) and decrease on the other side of this
point. Since $G>0$, the group velocity $v_{_g}$ increases/decreases with
increasing/decreasing wave number $k$. This means that energy of plasma waves
flows out of the minimum amplitude region in both directions towards regions of
higher and higher  amplitudes. As a result, the energy of plasma waves gets
packed into small regions where the amplitude grows even larger and the plasma
density decreases, forming a low density cavities. The effective force which
sweeps plasma particles out of the cavity is called ponderomotive force or the
Miller force \citep{gm58}, is just a measure of difference of the
high-frequency electromagnetic pressure between regions of high and low
amplitudes of plasma waves. This force is sensitive to mass and charge but
insensitive to the sign of charge of plasma particles. In the laboratory plasma
the ponderomotive force causes the effective charge separation due to huge
difference in the inertia of ions and electrons \citep[e.g.][]{sag79}. There
have been many laboratory experiments confirming existence of the Miller force
in an electron-ion plasma \citep[see also][for evidence of charge separation in
the Earth ionosphere]{petv76}.

\vskip
12pt~\scalebox{0.5}{\includegraphics*{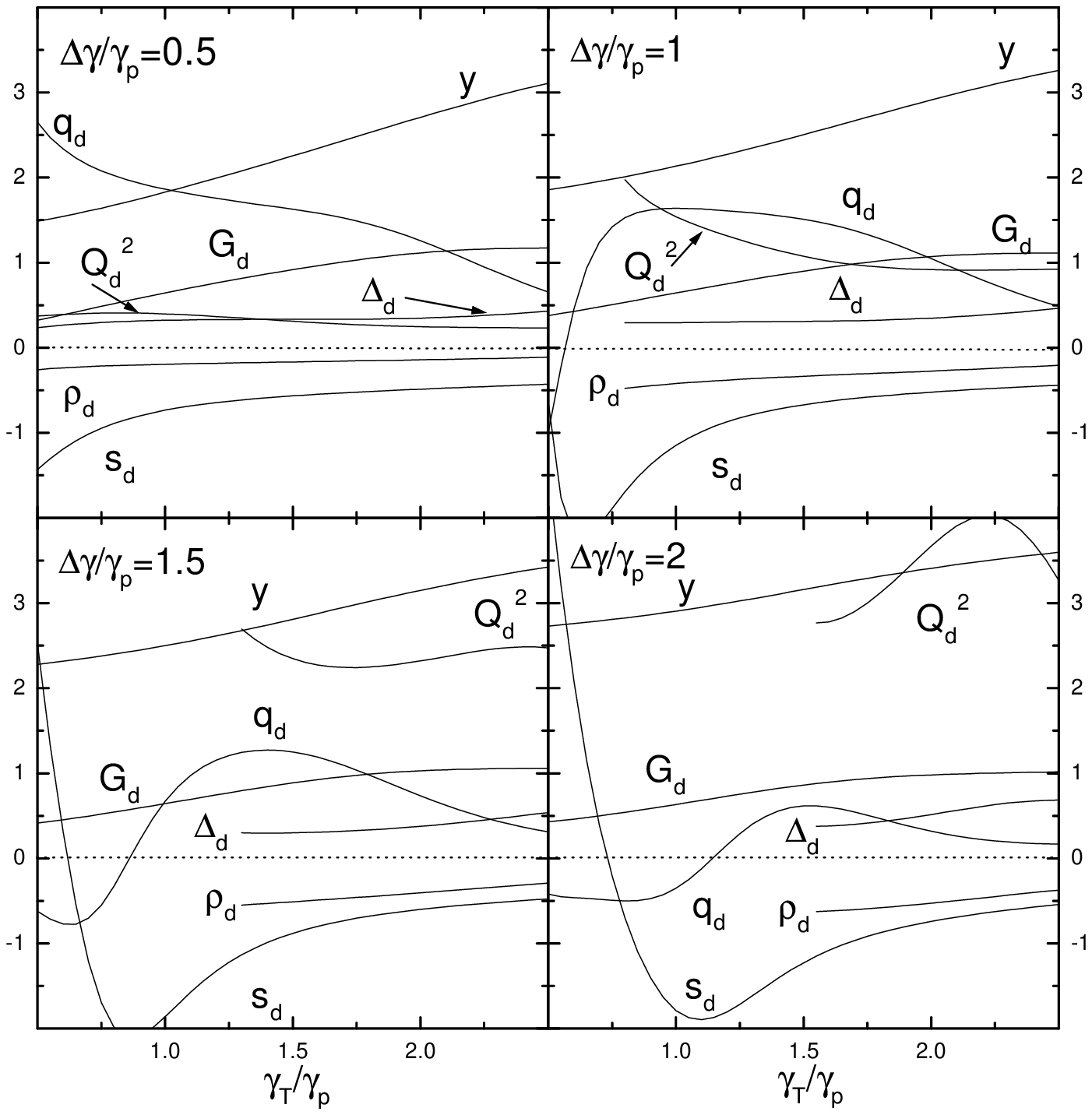}}~\figcaption[MGPFIG1.EPS]{Values of different parameters (see Appendix C for explanations and definitions) versus thermal spread defined as $\gamma_{_T}/\gamma_{_p}$, where $\gamma_{_T}\thicksim p_{_T}$ is a Lorentz factor describing a degree of plasma thermalization and $\gamma_{_p}$ is average Lorentz factor of ambient plasma particles, for four different values of relative shift of electron and positron distribution functions $\Delta\gamma/\gamma_{_p}=0.5, 1.0, 1.5, 2.0$, respectively. \label{fig1}}~\vskip 2pt

Let us then examine an influence of the Miller force on the electron-positron
pair plasma. If the distributions of electrons and positrons are identical,
then the Miller force affects electrons and positrons in the same rate, and
effective charge separation will not occur. This can be directly noticed from
equation (\ref{Def_rho_1}). However, in the pulsar magnetospheric plasma
$f_e\neq f_p$. As demonstrated by \citet{cr77a} the difference between $f_e$
and $f_p$ is a result of variation of the product ${\mathbf\Omega}\cdot{\mathbf
B}({\mathbf r})$ along a flux tube of dipolar magnetic field lines. Since the
numbers of electrons and positrons are equal, therefore the difference is in
the average Lorentz factors of pair plasma components. One can show that
$\Delta\gamma/\gamma_{_p}\simeq\Delta\sigma\gamma^3_{_p}/\gamma_{_b}$
\citep[e.g.][]{am98}, where $\Delta\gamma$ is the difference of average Lorentz
factors of electrons and positrons, and $\Delta\sigma=\sigma/\sigma_{_0}-1$ is
the normalised difference of the opening angles
$\sigma=\arccos\left({\mathbf\Omega}\cdot{\mathbf B}({\mathbf r})/(\Omega
B)\right)$ between the point under consideration ($\sigma$) and the stellar
surface ($\sigma_{_0}$) along a given dipolar field line (see Appendix
\ref{Defs} for definitions of $\gamma_{_p}$ and $\gamma_{_b}$). For a typical
pulsar at altitudes of about $50$ stellar radii $\delta\sigma\sim0.5$. Thus for
canonical values of Lorentz factors $\gamma_{p}\sim100$ and
$\gamma_{b}\sim10^6$ we have $\Delta\gamma/\gamma_{p}\sim0.5$, which is just
enough to cause significant charge separation due to effective relativistic
mass difference between electrons and positrons (see Fig.~1 for the parameter
$Q_{_d}$ appearing in eq.[\ref{L1}]). If the surface magnetic field is
non-dipolar, then $\Delta\gamma/\gamma_p$ can be even larger.

If a difference between the electron and positron distribution functions is
small (that is $\Delta\gamma/\gamma\thicksim 1$), the dispersion of linear
waves and the coefficients of the nonlinear Schr\"{o}dinger equation remain
almost unchanged with respect to the well-known case for which $\Delta\gamma=0$
\citep{mp80a,mp84}. Full details of derivation of the nonlinear Schr\"{o}dinger
equation for electrostatic waves in the relativistic electron-positron plasma
associated with succession of spark-generated clouds are presented in the
Appendix A. Below we outline the main results and discuss their implications
for effective mechanism of pulsar radio emission.

\vskip
12pt~\scalebox{0.5}{\includegraphics*{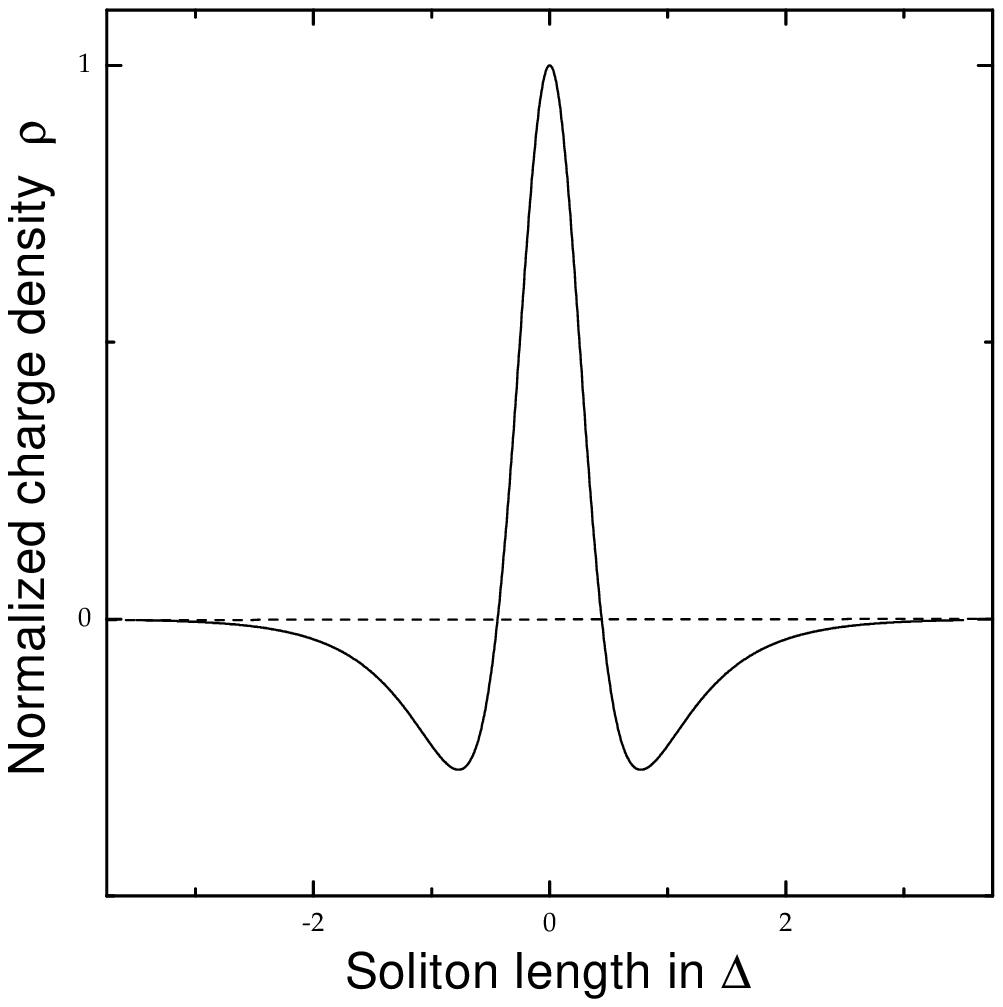}}~\figcaption[MGPFIG2.EPS]{Normalized charge density $\rho$ (eq.[\ref{Eq8}]) versus length in units of $\Delta$ (eq.[\ref{Delta}]) of a spark-associated soliton (see Fig. 3 for the model of charge distribution). \label{fig2}}~\vskip 2pt

As argued above, the soliton charge separation due to relative motion of
electrons and positrons is supported by the pondemotorive or so called the
Miller force \citep{gm58}.  In the LFR (Appendix A) the value of the charge
density contrast associated with solitons can be determined by means of
evaluating integrals in (\ref{Def_rho_1}) and substituting into it equation
(\ref{solution}). The expression for slowly varying charge density inside the
soliton has the form
\begin{equation}
\rho \thicksim n_{p}~e~\chi^{2}~\frac{\cosh ^{2}\zeta
-2}{\cosh^{3}\zeta}\rho_{d} \qquad
\rm{cm^{-{\frac{3}{2}}}~g^{\frac{1}{2}}~s^{-1}},
\label{Eq8}
\end{equation}
where $e$ is a fundamental charge, $n_p$ is a number density of unperturbed
ambient plasma, $\rho_{d}$ is dimensionless parameter about $\thicksim 0.3$
(Fig.\ref{fig1}),
\begin{equation}
\zeta =\frac{z-v_{g}t}{\Delta },
\label{Eq9}
\end{equation}
$\chi $ is defined by equation (\ref{Dim_amp}) as a ratio of Langmuir waves and
plasma energy densities, and
\begin{equation}
\Delta \approx \gamma_{_0}^{-1}~K_{m}^{-1}
\thicksim 36~\gamma_{_2}^{0.5}~\varkappa^{-0.5}~
R_{_{50}}^{1.5}\Delta_{d}~\chi^{-0.5}~P^{0.25}~\dot{P}_{_{-15}}^{-0.25}
\qquad \rm{cm}
\label{Delta}
\end{equation}
is the characteristic soliton length scale, i.e. its longitudinal (along the
magnetic field lines) dimension in LFR, where $\gamma_{_0}$ is a Lorentz factor
of relative motion of WFR and LFR (Appendix A), $K_{m}$ is the wave vector of
low frequency perturbation (\ref{ka}), and $\Delta_{d}$ is a dimensionless
parameter shown in  Figure \ref{fig1}. As one can see, the range of
characteristic soliton dimension $\Delta$ is between $10$ and $100$ cm in the
LFR, so they should be capable of emitting coherent curvature radiation at
radio wavelengths.

Figure \ref{fig2} presents schematically a charge distribution corresponding to
equation (\ref{Eq8}). This kind of charge distribution can be modeled as a
system of three charged bunches (Fig.\ref{fig3}) coupled to each other and
moving along the circular trajectory with a radius of curvature $r_c={\cal R}$.
The value of the central charge is $Q$, but the whole system is neutral as each
of the two side charges has value equal to $-\frac{1}{2}Q$ (an estimate of $Q$
will be given later in the paper). The ratio of the soliton charge density
(\ref{Eq8}) and Goldreich-Julian charge density $\rho_{_{GJ}}$ can be estimated
as
\begin{equation}
\frac{\rho}{\rho_{_{GJ}}}\thicksim \varkappa~\chi^2~\rho_d,
\label{Eq11}
\end{equation}
where $\varkappa\thicksim 10^4$, $\chi\thicksim 0.1$ (see discussion below
eq.[\ref{Dim_amp}]) and $\rho_d \thicksim 0.3$ (see Figure \ref{fig1}). Thus,
the soliton charge density is about $10$ times the Goldreich-Julian
corotational value $\rho_{_{GJ}}=e n_{_{GJ}}$ (eq.[\ref{nGJ}]) or about
$10^{-3}$ of the secondary (Sturrock multiplicated) ambient plasma charge
density $n_p$ (see eq.[\ref{Eq8}]).

\vskip
12pt~\scalebox{0.5}{\includegraphics*{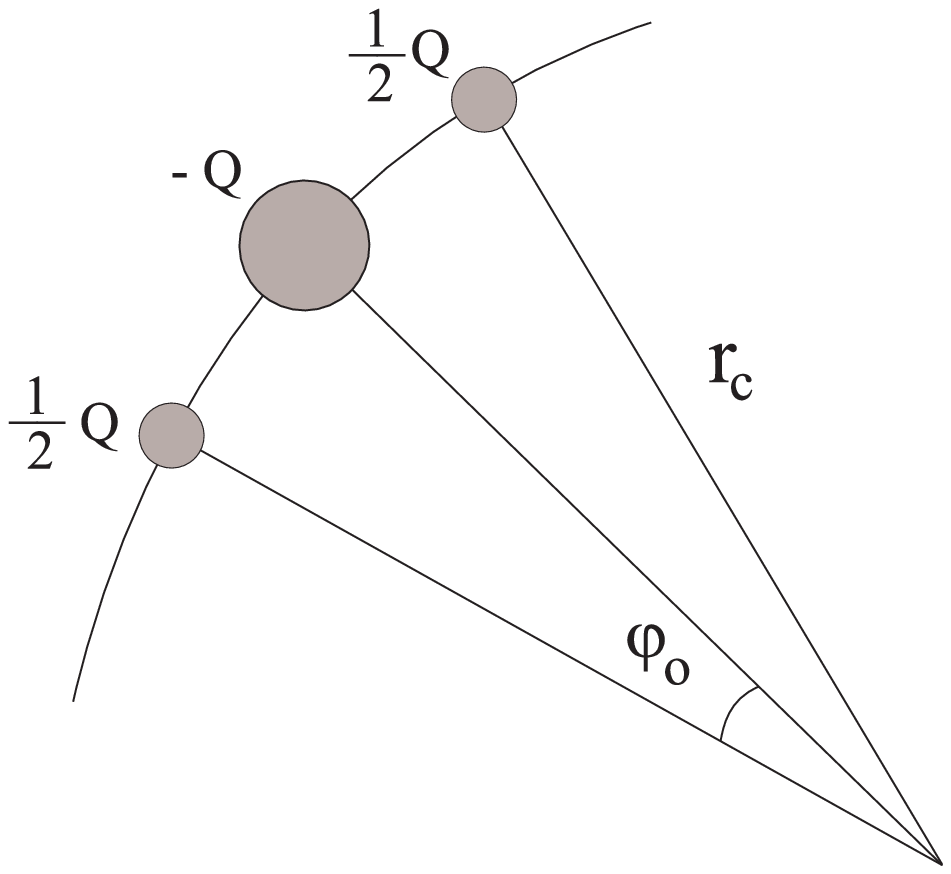}}~\figcaption[MGPFIG3.EPS]{A
simple model for the soliton with charge density shown in Fig. 2. The
characteristic soliton length $\Delta=\varphi_{o}r_c$, where $r_c$ is a local
radius of curvature of the soliton trajectory. \label{fig3}}~\vskip 2pt

Now let us examine the coherent curvature radiation of a centimeters long (in
LFF) soliton bunches. The equation (\ref{dInoc}) for the infinitesimal
radiation intensity $dI_{n}$ differs from the well-known equation for single
particle curvature radiation intensity only by a term $\left[ 1-\cos\left(
n\varphi_{o}\right)\right]^{2}$, which does not depend on the solid angle
$\sigma $. According to detailed calculations presented in Appendix B, the
spectral power of our soliton is
\begin{equation}
I_{\omega }=\frac{Q^{2}}{c}\omega_{_0}F\left( \frac{\omega }{\omega_{c}}
\right) \left[1-\cos\left( a\frac{\omega }{\omega_{c}}\right) \right]^{2}=
\frac{Q^{2}}{c}\omega_{_0}F\left( \frac{\omega }{\omega_{c}}
\right) \left[1-\cos\left( \frac{2\pi }{c}\frac{\Delta}{\lambda}\right)
\right]^{2},
\label{Eq12}
\end{equation}
where
$\omega_{c}=\frac{3}{2}\gamma_{_0}^{3}\omega_{_0}\approx\gamma_{_0}^3c/{\cal
R}$ is the characteristic frequency of single-particle curvature radiation,
$\lambda$ is an emitted wavelength, the function
\begin{equation}
F(x)=x \int\limits_{x}^{\infty }K_{\frac{5}{3}}(x)dx ,
\label{Eq13}
\end{equation}
where $K_{\frac{5}{3}}(x)$ is the modified Bessel function and the parameter
$a=\frac{\Delta}{\mathcal R}\gamma_{_0}^3\ll\frac{\pi}{2}$ (Appendix B) or
\begin{equation}
a \approx
8\times10^{-2}~\gamma_{_2}^{3.5}~\varkappa_{4}^{-0.5}R_{_{50}}~y^3~\Delta_{d}~\chi^{-0.5}~P^{-0.25}~{\dot{P}_{_{-15}}}^{-0.25}.
\label{a}
\end{equation}
Here $y=\gamma_{_0}/\gamma_{_p}$ is a dimensionless parameter describing the
ratio of the Lorentz factors of solitons and bulk plasma particles in the radio
emission region which is about unity for typical pulsar parameters.

To calculate the spectral power $I_{\omega}$ (eq.[\ref{Eq12}]) we have to find
the soliton charge $Q$ by integrating the charge density $\rho$
(eq.[\ref{Eq8}]) over the soliton characteristic volume ${\cal V}\thicksim
S_{\perp}\Delta$. Here $S_{\perp}$ is a soliton cross section, which we can
estimate assuming that the perpendicular (with respect the magnetic field) size
of the spark-associated plasma clouds at the stellar surface is about the gap
height $h \simeq 5\times 10^3 P^{3/7}$ cm \citep{gs00} and increases with a
radial distance as $r^3$. Therefore
\begin{equation}
S_{\perp }\thicksim 3\times 10^{12}~R_{_{50}}^{3}~P^{\frac{2}{7}}~
\dot{P}_{_{-15}}^{-\frac{4}{7}}\qquad\rm{cm^2},
\label{Eq15}
\end{equation}
and
\begin{equation}
{\cal V}\thicksim
10^{14}~\gamma_{_2}^{0.5}\varkappa_{4}^{-0.5}~R_{_{50}}^{4.5}~
\Delta_d~\chi^{-0.5}~P^{0.54}~\dot{P}_{_{-15}}^{-0.82}\qquad\rm{cm^3},
\label{Eq16}
\end{equation}
which is about $10^{14}~\rm{cm^3}$. Thus, a soliton charge
$Q\simeq10\rho_{_{GJ}}{\cal V}$, where $\rho_{_{GJ}}=en_{_{GJ}}\thicksim e10^6
{}~\rm{cm^{-3}}$ at the altitudes of about $50R$ (eq.[\ref{nGJ}]), which gives
$Q\simeq e10^{21}$.
Therefore,
\begin{equation}
I_{\omega}\thicksim 1.4\times 10^{18}\gamma_{_2}
\varkappa_{4}R_{_{50}}^{2.5}\chi^{3}P^{-0.43}
\dot{P}_{_{-15}}^{-0.64}Q_{d}^{2}~F\left(\frac{\omega}{\omega_{c}}\right)
\left[1-\cos\left(a\frac{\omega}{\omega_{c}}\right)\right]^2 ~
\rm{erg~rad^{-1}}.
\label{Iom}
\end{equation}
Consequently, the power radiated by one soliton $L_1=\int
I_{\omega}d\omega\simeq 4~\nu_c~I_{\nu}$, where $\nu_c\sim\gamma_{_0}^3c/{\cal
R}$ (see Fig.\ref{fig4}) is the characteristic frequency of curvature
radiation, is about
\begin{equation}
L_{1}\thicksim 10^{22}~ \gamma_{_2}^4 ~\varkappa_4~R_{_{50}}^{2}
P^{-0.93}~\dot{P}_{_{-15}}^{-0.64}
\left[
\frac{y^3~\chi^3~Q_{d}^{2}~I_{o}\left(a\right)}{1.2\times10^{-4}}
\right]
\qquad \rm{erg~s^{-1}},
\label{L1}
\end{equation}
where $Q_{d}$ and $I_{o}\left(a\right) =\int F(x)\left[1-\cos(ax)\right]^2dx$
depend on the parameters of the plasma and their values for different cases are
shown in Figures \ref{fig1} and \ref{fig5}, respectively. This power is
radiated mainly in the narrow frequency band around
\begin{equation}
\nu_{m}\thicksim 4.4\times
10^{7}~\gamma_{_2}^{3}~y^3~R_{_{50}}^{-0.5}~P^{-0.5}\qquad \rm{Hz},
\label{nm}
\end{equation}
which is about $4$ times more than $\nu_c=\omega_c/2\pi$ (see Fig.\ref{fig4}).
Apparently, for $R_{_{50}}\thicksim 2$, $\gamma_{_{2}}\thicksim 1$ and $y
\thicksim 2.3$ this maximum frequency is close to $400$ MHz, around which
pulsars appear brightest.

\vskip
12pt~\scalebox{0.5}{\includegraphics*{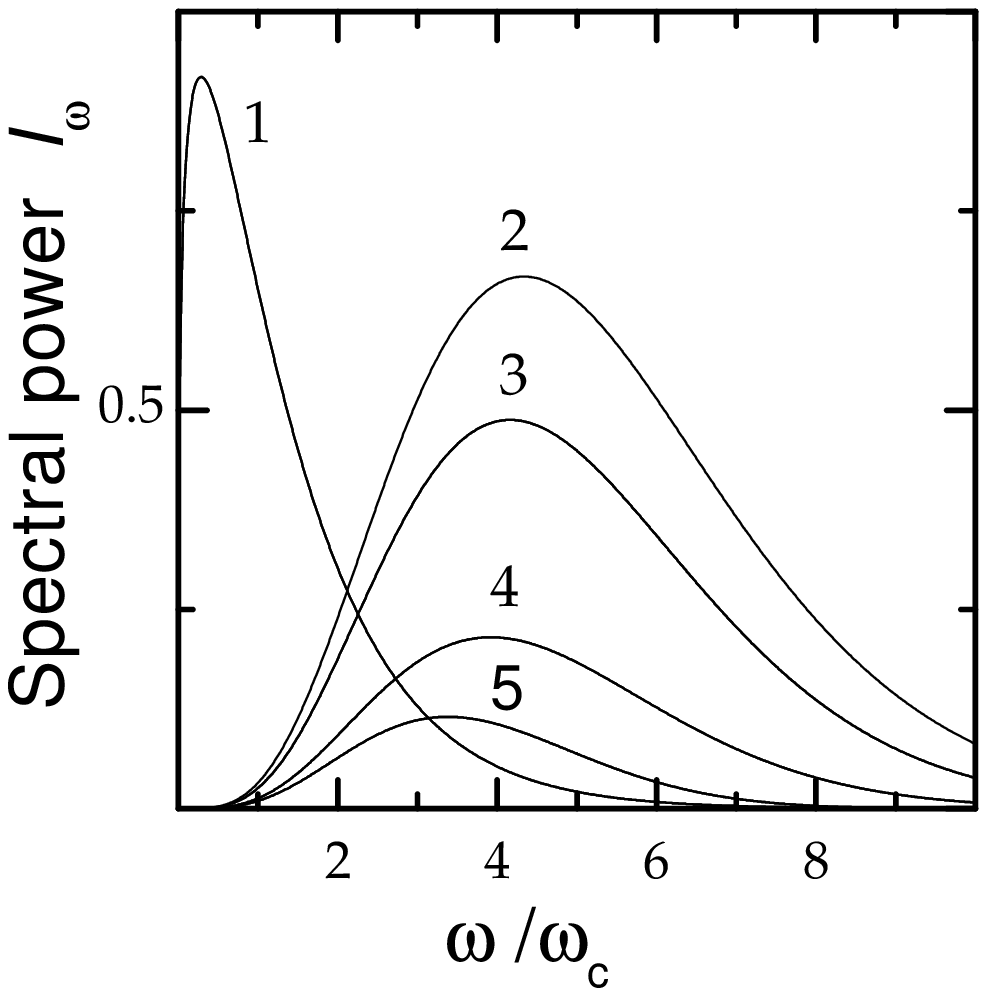}}~\figcaption[MGPFIG4.EPS]{Normalized spectral power $I_{\omega}$ versus $\omega/\omega_c$, where $\omega$ is the frequency of curvature radiation and $\omega_c=1.5\gamma_{_0}^3c/r_c$ (Appendix C) for: (1) single particle curvature radiation, and soliton curvature radiation with (2) $a=0.1$, (3) $a=0.2$, (4) $a=0.3$ and (5) $a=0.5$, where $a$ is described in equation (\ref{a}). The cases (1), (2), (3) and (4) are multiplied for clarity of presentation by a factor of $2000$, $100$, $10$ and $1$, respectively (see eqs.[12-18]) Notice that the frequency $\nu_m$ for which spectral power reaches maximum is about 4 times higher for solitons than for single particle curvature radiation. \label{fig4}}~\vskip 2pt

The total number, $N_t$, of solitons contributing to pulsar radiation at any
instant can be calculated as the number of sparks  $N_{sp}$ on the polar cap
multiplied by the number of solitons $N_{sl}$ associated with each spark.
Since, as it is clear from figures 2 and 3, $N_{sl} \simeq \Delta r/(10\Delta)
$, where $\Delta r\thicksim R_{_{50}}\times 50R$ and $\Delta$ is a soliton
length scale expressed in equation (\ref{Delta}). Thus
\begin{equation}
N_{sl}\thicksim 1.4\times 10^{5}~ \gamma_{_2}^{-0.5}~\varkappa_4^{0.5}~
R_{_{50}}^{-0.5}~\Delta_d^{-1}~\chi^{0.5}~P^{-0.25}~\dot{P}_{_{-15}}^{0.25}.
\label{Nsl}
\end{equation}
The number of sparks on the polar cap with a radius $r_p\thicksim
10^4P^{-0.5}$~cm is approximately equal to $(r_{_{p}}/h)^2$, where $r_{_{p}}$
is the polar cap radius and $h\thicksim 5\times 10^3P^{3/7}$~cm is the polar
gap height \citep[for details see][]{gs00}. Thus
\begin{equation}
N_{sp}\thicksim 25~P^{-1.3}~\dot{P}_{_{-15}}^{0.57}
\label{Nsp}
\end{equation}
and
\begin{equation}
N_{t}\thicksim N_{sp}\times N_{sl}\thicksim 3.5\times 10^6~
\gamma_{_2}^{-0.5}~\varkappa_4^{0.5}~
R_{_{50}}^{-0.5}~\Delta_d^{-1}~\chi^{0.5}~
P^{-1.54}~ \dot{P}_{_{-15}}^{0.82}.
\label{Nt}
\end{equation}
Consequently, the total power $L_t=L_1~N_t$ radiated by a pulsar can be
estimated as
\begin{equation}
L_{t}\thicksim 10^{28}~ \gamma_{_2}^{3.5}~\varkappa_4^{1.5}~R_{_{50}}^{1.5}
P^{-2.47}~\dot{P}_{_{-15}}^{0.18}
\left[
6 y^3~\chi^{3.5}~Q_{d}^{2}~I_{o}\left(a\right)\Delta_d^{-1}\times10^{4}
\right]
\qquad \rm{erg~s^{-1}}.
\label{Lt}
\end{equation}
Since the expression in square brackets is of the order of unity, it is clear
from equation (\ref{Lt}) that the power radiated by the spark-associated
solitons can easily explain the observed radio emission of pulsars. In fact,
for a typical pulsars the apparent luminosity is in the range $10^{25} -
10^{29}~\rm{erg~s^{-1}}$ (see also Table~1).

\section{Discussion}

In this paper we propose a new, self-consistent theory of pulsar radio emission
based on the modified non-stationary sparking model of \citet{rs75}. As argued
by Gil \& Sendyk in the accompanying Paper I, the polar cap with radius $r_p
\thicksim 10^4P^{-1/2}$~cm is populated with about $(r_p/h)^2$ sparks of a
characteristic dimension approximately equal to the polar gap height $h
\thicksim 5\times 10^3P^{3/7}$~cm. Each repeatable spark delivers to the open
magnetosphere a sequence of $e^{-}e^{+}$ clouds, flowing orderly along dipolar
magnetic field lines. Overlapping of particles with different momenta from
consecutive sparks leads to a two-stream instability, which triggers a strong
electrostatic Langmuir waves at the altitudes of about $50$ stellar radii. This
is the only known effective instability which can develop at low altitudes,
where the narrow-band pulsar radiation originates \citep[e.g.,][]{kg97,kg98}.
These oscillations with a characteristic frequency of about $100$~GHz, are
modulationally unstable and their nonlinear evolution results in the formation
of `bunch-like' charged solitons. A characteristic soliton length along the
magnetic field lines is about $30$~cm as viewed by a distant observer, so they
are capable of emitting coherent curvature radiation at radio wavelengths. A
perpendicular cross-section of each soliton at radiation altitudes results from
a dipolar spread of a plasma clouds with a characteristic dimension of about $h
\thicksim 50$~m near the neutron star surface. The net soliton charge $Q$ is
about $10^{21}$ fundamental charges $e$, contained within a volume of about
$10^{14}~ \rm{cm^3}$. For a typical pulsar ($P\sim 1s, \dot{P}\sim 10^{-15}$),
there are about $10^5$ solitons associated with each of about 25 sparks
operating on the polar cap at any instant. Since one soliton moving along
dipolar field lines with a Lorentz factor $\gamma$ of the order of $100$
generates a power of about $10^{21}~ \rm{erg~s^{-1}}$, then a total power of
typical pulsar can be estimated as about $10^{27}~ \rm{erg~s^{-1}}$. The degree
of coherence for radio wavelengths is highest for smallest solitons. On the
other hand, the net charge increases with the soliton size. This implies that
the pulsar radiation should have a maximum intensity at some intermediate
wavelength between 20 and 70 centimeters. This is really observed in most
pulsars \citep[e.g.][]{mal94}.

One of the reasons for such a huge net charge is that the soliton volume is
large. Although only 30 cm in length along field line, it spans the entire
transverse of the spark-associated plasma column. We make an implicit
assumption that the soliton is transversely stable, at least over a time
interval during which its curvature radiation is emitted towards an observer.
Wheatherall (1997, 1998) has shown that an extremely strong plasma wave
turbulence originating within the two-stream instability suffers rapid collapse
transverse to the magnetic field. However, this effect is not relevant to our
theory, since soliton amplitudes are limited by the non-linear Landau damping.

As far as millisecond pulsars are concerned, we cannot perform luminosity
calculations for the following reason: the surface magnetic field inferred from
$P$ and $\dot{P}$ values is very small, typically $B_{12}^s\sim 10^{-4}$. This
implies very large canonical gap height $h\sim 5\times 10^3{\cal
R}_6^{2/7}B_{12}^{-4/7}P^{3/7}~{\rm cm}$, typically close to $10^5$~cm.
Consequently, the spark-generated two-stream instability region would be
located outside the light-cylinder and our theory can not be applied in the
present form to the millisecond pulsars. However, it is quite possible that the
surface field of millisecond pulsars is much stronger than the value inferred
from the magnetic dipole braking law \citep[see arguments by][and references
therein]{cgz98,cz99,gm99}. If this is the case, then everything scales properly
and millisecond pulsars are not different from normal pulsars within the
framework of our theory. We will present detailed intensity calculations for
millisecond pulsars in a separate paper.

In our approach, contrary to those taken in previous studies, the problem of
the formation of charged bunches is resolved automatically and
self-consistently. The key feature of our picture is an existence of
short-living ($\thicksim 10^{-5}$ s) sparks with both a characteristic
dimension and typical distance between them equal approximately to the polar
gap height $h\sim 5\times 10^3P^{2/7}$~cm \citep{gs00}. As a result of their
repeatable operation in approximately the same place, a two stream instability
develops at low altitudes below 10\% of the light cylinder radius, which
generates high-frequency ($\nu \sim 100$ MHz) Langmuir electrostatic waves
within the spark-associated plasma columns.

The soliton formation due to nonlinear development of strong Langmuir
electrostatic waves is warranted by the hydrodynamical type of the linear
instability, in which half of the energy of streaming charges can be
transferred to the plasma turbulence. Because of the relative motion of centers
of mass of electrons and positrons, the pondemotorive force acts on them
differently, redistributing charges over the soliton volume. This results in a
net soliton charge, consisting in fact from a system of three coupled localized
charges. The soliton curvature radiation is much less effective than the
curvature radiation of single small bunch with the same charge. However, since
the soliton charge is huge, its curvature radiation is powerful enough to
account for the observed pulsar radio emission. It is worth emphasizing that
this radiation is supported by the kinetic energy of secondary e$^-$e$^+$
plasma with $\gamma_{_p}\sim 100$, created via magnetic pair production by the
primary beam with $\gamma_{_b}\sim 10^6$ produced by the accelerating potential
drop near the polar cap. It should be realized that a significant fraction of
kinetic energy produced by sparks is radiated away in the form of soliton
curvature radiation at radio wavelengths. In fact, the maximum kinetic
luminosity associated with $N_{sp}$ sparks (eq.[21]) is $L_m\simeq
N_{sp}\,\dot{N}_s\,e\,f\,V$, where $\dot{N}_s\simeq
7\times10^{10}{P}^{1/2}_{-15}P^{-1/2}\pi h^2c$ is the particle energy flux
associated with a single spark. Here $1>f>{\cal F}$, where ${\cal F}=0.1{\cal
R}_6^{-5/21}B_{12}^{1/7}P^{1/7}$ is the filling factor \citep[see][]{gs00} such
that the actual accelerating potential of the developing  spark is $\Delta
V=f\,V$, where $V$ is described by the maximum potential drop expressed in
equation (1). Therefore, $L_m=2.4\times 10^{30}f{\cal
R}_6^{8/7}B_{12}^{-9/7}\left(\dot{P}_{-15}/P\right)^{15/14}~{\rm erg~s^{-1}}$,
and thus for ${\cal R}^{8/7}B_{12}^{-9/7}\sim 0.1$ and $f\sim 0.25$, we can
write approximately
\begin{equation}
L_m\approx 5\times 10^{28}\dot{P}_{-15}P^{-1}~~ {\rm erg~s^{-1}},
\label{Eq24}
\end{equation}
which is about an observable pulsar total radio luminosity.

In Table 1 we present results of the luminosity calculations from equation (23)
for a number of pulsars with different values of period $P$ and period
derivative $\dot{P}=10^{-15}\dot{P}_{-15}$. As one can see, it is easy to
obtain the total luminosity $L_t$ close to the observed luminosity
$L_{_R}\thicksim 3.5\times10^{25+x}$, where $x=\mathrm{Log}~(L)$ is taken from
Table 4 in the pulsar catalog \citep{tml93}, for a narrow range of parameters
$\gamma_{_2}=\gamma_{_p}/100$ and $y=\gamma_{_0}/\gamma_{_p}$ (Fig.\ref{fig1}).
This means that the pulsar luminosity $L_R\sim L_t$ is determined mainly by the
values of $P$ and $\dot{P}$, similarly to the morphological properties of
single pulses and average profiles \citep{gs00}. The fraction $f=L_R/L_{sp}$,
where $L_{sp}=3.8\times 10^{31}\dot{P}_{-15}P^{-3}$~erg/s is the pulsar
spindown luminosity \citep{tml93}, is a small number between $10^{-9} -
10^{-3}$, increasing towards longer periods, as observed. This is easy to
understand bearing in mind that the soliton pulsar radiation is supported by
the kinetic energy generated by the accelerating potential drop within the
polar gap. In fact, if a significant fraction (say 30\%) of spark maximum
luminosity $L_m$ (eqs.[\ref{Lt}] and [\ref{Eq24}]) is converted to the soliton
curvature radio emission, that is $L_R\sim L_t\sim 0.3\,L_m$, then
$L_R/L_{sp}\sim 0.3\times 10^{-4}\,P^2$. This ratio is about $3\times 10^{-8}$
for the Crab pulsar, $2\times 10^{-7}$ for Vela pulsar, $3\times 10^{-5}$ for
one second pulsar, $4\times 10^{-4}$ for long period (3.75 s) pulsar and
$2\times 10^{-3}$ for longest period (8.5 s) pulsar J2144-3933 (Young et al.
1999), in good agreement with data presented in Table 1.

In the framework of the soliton model presented in this paper, the frequency of
radiated waves is much smaller than the characteristic frequency of the ambient
plasma. For small angles between the wave vector and the external magnetic
field, which is the case of the curvature radiation, spectra of these waves is
$\omega=kc(1-\delta)$, where
$\delta\propto\left(\omega_{_p}/\omega_{_B}\right)^2$ is negligibly small in
the inner magnetosphere \citep[e.g.,][]{lmmp86}. Also \citet{kmm91} considered
wave propagation in relativistic electron-positron pulsar magnetospheric plasma
with $\Delta\gamma=\gamma_{+}-\gamma_{-}\neq0$ and showed that, in the case of
propagation nearly along curved magnetic field lines there exist two
elliptically polarised almost electromagnetic waves (with very small potential
components). Therefore the curvature radiation at frequencies well below
$\omega_{_l}$ generated by relativistic soliton embedded in a surrounding
plasma do have a dominating electromagnetic feature and can propagate through
plasma, like in a vacuum.

Since the characteristic longitudinal dimension of our solitons is smaller than
the emitted radio wavelengths, the polarization properties should be the same
as in the case of single-particle curvature radiation modulated by the
spark-associated envelope function \citep{gsn90a}, that is: (i) relatively high
linear polarization with a position angle swing across the plane of the source
motion and moderate circular polarization with sense reversals in the same
plane \citep{mich87,gsn90b,gkz93,gan97,gan99}; (ii) the core components should
differ from the conal components polarization-wise as a result of subpulse
drift and/or scatter in conal parts of the mean profile \citep{gkms95,gs00}. In
particular, the position angle in the core components should swing faster than
predicted by the rotating vector model \citep{rc69}, while in conal components
the mean position angle curve should follow the rotating vector model more
closely. On the other hand the circular polarization should typically change
sense near the intensity maximum of core components, while in the conal
components the circular position should be rather weak and mostly of one sense.
All these specific properties are really observed
\citep{r83,lm88,r90,r93,gl95}.

\acknowledgments

This paper is supported in part by the KBN Grant 2~P03D~015~12 of the Polish
State Committee for Scientific Research. GIM was also supported by the INTAS
Grant 96-0154. We thank D. Lorimer, K.S. Cheng and D. Mitra for helpful
discussion and E. Gil for technical assistance.

\appendix

\section{Nonlinear Schr\"{o}dinger equation\label{A}}

The nonlinear Schr\"{o}dinger equation is accepted as the fundamental equation
to describe the nonlinear wave modulation in various kinds of the dispersive
media \citep{kk69,ty69,zs72}. Usually when studying the nonlinear evolution of
the Langmuir waves, the low frequency perturbation due to ion-acoustic waves is
considered \citep[e.g.,][]{zs72}. But in the case of electron-positron plasma
there is no such low frequency perturbation. An important effect associated
with the resonance effects of particles moving at the  group velocity was
investigated by \citet{it73} and they derived a nonlinear Schr\"{o}dinger
equation taking into account the nonlinear Landau damping. Contributions of the
resonance particles at the group velocity of the wave modifies drastically the
structure of the nonlinear Schr\"{o}dinger equation in two respects. One is the
appearance of a nonlocal-nonlinear term, which indeed arises from the nonlinear
Landau damping process associated with the resonance at the group velocity of
the wave. The other is the fact that the coupling coefficient of the
local-nonlinear term is also modified by the contributions of these resonant
particles. The relativistic case of this problem (when the plasma particle
velocities are near the speed of light) was studied by \citet{mp80a}. They
found that the resonant interaction between the nonlinear beatings and
particles is of the primary importance in the case of electron-positron plasma
\citep{mp80b}.

Let us introduce two reference systems: the first system is connected with the
neutron star and we will call it a {\it laboratory frame of reference} (LFR),
the second one is moving with respect to the LFR with the velocity equal the
group velocity of the of the Langmuir waves. We will call this system the {\it
wave frame of reference} (WFR). We can study a one-dimensional case, which is
correct for the pulsar magnetosphere as long as the magnetic field is strong
enough to control the plasma motion. Let us assume that the external magnetic
field is directed along the $z$-axis and that all perturbations are directed
along the magnetic field. The $z$ and time $t$-coordinates in the WFR is
connected with LFR by the following formulas
\begin{equation}
z^{\prime }=\gamma_{_0}\left( z-v_{g}t\right) ,\qquad t^{\prime
}=\gamma_{_0}\left(t-\frac{v_{g}}{c^{2}}z\right).
\label{coord}
\end{equation}
Here $v_{g}=\partial\omega_{_l}/\partial k_{_l}$ is the group velocity of
Langmuir waves and $\gamma_{_0}=\left( 1-v_{g}^{2}/c^{2}\right) ^{-1/2}$. Thus,
$v_{g}$ and $\gamma_{_0}$ are directly related with solitons propagating in an
ambient secondary plasma with an average Lorentz factor $\gamma_{_p}$ (Table
1). In order to describe a nonlinear behavior of the wave packet we introduce
the following representation of a distribution function and fields
\begin{equation}
F=F^{(0)}+\sum\limits_{m=-\infty }^{\infty }\sum\limits_{n=1}^{\infty
}\varepsilon ^{n}F_{m}^{(n)}(\xi ,\tau )\exp \left( il\left( k_{_l}^{\prime
}z^{\prime }-\omega_{_l}^{\prime }t^{\prime }\right) \right),
\label{sum}
\end{equation}
where $k_{_l}^{\prime }$ is a component of the wave vector along the magnetic
field, $\omega_{_l}^{\prime }$ is the frequency of plasma waves in the WFR,
$\varepsilon $ is a small parameter of the perturbation theory and
\begin{equation}
\xi =\varepsilon z^{\prime },\qquad \tau =\varepsilon ^{2}t^{\prime}.
\label{eps_eps_sq}
\end{equation}
This procedure separates slow and fast perturbations of plasma and wave
parameters. Consequently, we are assuming that $\partial /\partial
\tau\rightarrow \varepsilon ^{2}$ and $\partial /\partial \xi \rightarrow
\varepsilon $. To get a nonlinear Schr\"{o}dinger equation it is sufficient and
enough to derive a current of third order. Therefore we keep the sum in
equation (\ref{sum}) for values: $n=1,2,3$\ and $m=\pm 1,\pm 2,\pm 3$.
Substituting equations (\ref{coord} - \ref{eps_eps_sq}) into the well-known
kinetic Vlasov equation as well as satisfying the Maxwell equations, we obtain
in the above approximation the so called nonlinear Schr\"{o}dinger equation
\begin{equation}
i\frac{\partial }{\partial \tau }E_{\parallel }^{(1)}+G\frac{\partial
^{2}}{\partial \xi ^{2}}E_{\parallel }^{(1)}+q\left| E_{\parallel
}^{(1)}\right|^{2}E_{\parallel }^{(1)}+s\frac{1}{\pi }\oint \frac{\left|
E_{\parallel}^{(1)}(\xi ^{\prime }, \tau )\right| ^{2}}{\xi -\xi ^{\prime
}}d\xi ^{\prime}E_{\parallel }^{(1)}=0.
\label{shro}
\end{equation}
This equation is written in the WFR, but of course the electric field
$E_{\parallel }^{(1)}$ does not change when the frames are moving along the
$z$-axis. The coefficients of the equation (\ref{shro}) are following:
\begin{equation}
G=\frac{1}{2}\frac{d^{2}\omega_{_l}^{\prime }}{dk_{_l}^{\prime 2}}=\frac{1}{2}
\gamma_{_0}^{3}\frac{d^{2}\omega_{_l}}{dk_{_l}^{2}},
\label{Def_G}
\end{equation}
\begin{equation}
q=-\frac{(\omega_{_l}-k_{_l}v_{g})}{2k_{_l}}\left\{ \left(
\frac{A^{2}}{6k_{_l}}+\frac{B}{2}\right) -k_{_l}\left( \frac{\left(
W^{2}-V^{2}\right) H+2WVU}{H^{2}-V^{2}}+C\right) \right\},
\label{Def_q}
\end{equation}
\begin{equation}
s=\frac{(\omega_{_l}-k_{_l}v_{g})}{2k_{_l}}\left(
\frac{\left(W^{2}-V^{2}\right) U-2WVH}{H^{2}+V^{2}}+D\right),
\label{Def_s}
\end{equation}
where
\begin{equation}
W=\sum\limits_{\alpha }\omega_{p\alpha }^{2}\oint \frac{1}{\left(
\omega_{_l}-k_{_l}v\right)^{2}}\frac{dv}{dp}\frac{1}{\left(
v-v_{g}\right)}\frac{df_{\alpha }}{dp}dp ~~,
\label{Def_W}
\end{equation}
\begin{equation}
V=\sum\limits_{\alpha }\omega_{p\alpha }^{2}\int \frac{1}{\left(
\omega_{_l}-k_{_l}v\right) ^{2}}\frac{dv}{dp}\frac{df_{\alpha }}{dp}\delta
\left(v-v_{g}\right) dp ~~,
\label{Def_V}
\end{equation}
\begin{equation}
A=\sum\limits_{\alpha }\omega_{p\alpha }^{2}\int \frac{1}{\left(
\omega_{_l}-k_{_l}v\right) }\frac{d}{dp}\left( \frac{1}{\left( \omega_{_l}
-k_{_l}v\right)}\frac{df_{\alpha }}{
dp}\right) dp ~~,
\label{Def_A}
\end{equation}
\begin{equation}
B=\sum\limits_{\alpha }e_{\alpha }\omega_{p\alpha }^{2}\int \frac{1}{\left(
\omega_{_l}-k_{_l}v\right) }\frac{d}{dp}\left\{ \frac{1}{\left( \omega
_{l}-k_{_l}v\right) }\frac{d}{dp}\left( \frac{1}{\left( \omega
_{l}-k_{_l}v\right) }\frac{df_{\alpha }}{dp}\right) \right\} dp ~~,
\label{Def_B}
\end{equation}
\begin{equation}
C=-\sum\limits_{\alpha }e_{\alpha }\omega_{p\alpha }^{2}\oint \frac{1}{
\left( \omega_{_l}-k_{_l}v\right) ^{2}}\frac{dv}{dp}\frac{1}{\left(
v-v_{g}\right) }\frac{d}{dp}\left( \frac{\left( v-v_{g}\right) }{\left(
\omega_{_l}-k_{_l}v\right) ^{2}}\frac{df_{\alpha }}{dp}\right) dp ~~,
\label{Def_C}
\end{equation}
\begin{equation}
D=\sum\limits_{\alpha }e_{\alpha }\omega_{p\alpha }^{2}\int \frac{1}{\left(
\omega_{_l}-k_{_l}v\right) ^{2}}\frac{dv}{dp}\delta \left( v-v_{g}\right)
\frac{d}{dp}\left( \frac{\left( v-v_{g}\right) }{\left( \omega
_{l}-k_{_l}v\right) ^{2}}\frac{df_{\alpha }}{dp}\right) dp ~~,
\label{Def_D}
\end{equation}
\begin{equation}
H=\sum\limits_{\alpha }\omega_{p\alpha }^{2}\oint \frac{1}{\left(
v-v_{g}\right) }\frac{df_{\alpha }}{dp}dp ~~,
\label{Def_H}
\end{equation}
\begin{equation}
U=\sum\limits_{\alpha }\omega_{p\alpha }^{2}\int \delta \left(
v-v_{g}\right) \frac{df_{\alpha }}{dp}dp ~.
\label{Def_U}
\end{equation}
Here $\alpha \ $defines the sort of particles, $\omega_{_l},k_{_l}$ and $v$ are
defined in the LFR. It should be mentioned that the equation (\ref{shro}) is
written in WFR, but all the values in the integrals are defined in LFR. In this
paper we  use dimensionless momentum normalized to $m_ec$.

Nonlinear evolution of the waves described by the equation (\ref{shro}) is well
known \citep[e.g.,][]{ist73}. The maximum growth rate $\Gamma_{m}$\ of the
modulational instability is defined as
\begin{equation}
\Gamma_{m}=\left( q^{2}+s^{2}\right) ^{\frac{1}{2}}\left| E_{\parallel
o}^{(1)}\right|^{2},
\label{gamma}
\end{equation}
and a corresponding wave vector of low frequency perturbation is
\begin{equation}
K_{m}=\left( \frac{q^{2}+s^{2}}{Gq}\left| E_{\parallel o}^{(1)}\right|
^{2}\right) ^{\frac{1}{2}}.
\label{ka}
\end{equation}
In the case when $\left| q\right| \gg \left| s\right| $ the equation
(\ref{shro}) has the following solution
\begin{equation}
E_{_{\parallel }}^{(1)}(\xi ,\tau )=E_{_{\parallel }o}^{(1)}\mathrm{sech}\left(
E_{_{\parallel }o}^{(1)}\sqrt{\frac{q}{2G}}\left( \xi -u\tau \right) \right)
\exp \left\{ i\left( \frac{u}{2G}\xi -\frac{u^{2}}{4G}\tau +\frac{1}{2}q\tau
\left( E_{_{\parallel }o}^{(1)}\right) ^{2}\right) \right\}.
\label{solution}
\end{equation}
The slowly-varying charge density associated with developing soliton is given
by the following formula
\begin{equation}
\rho^{\prime }=\frac{\sum\limits_{\alpha }e_{_\alpha}\omega_{p \alpha
}^{2}\oint \frac{1}{\left( v-v_{g}\right)}\frac{d}{dp}
\left( \frac{\left( v-v_{g}\right)}{\left( \omega_{_l}-k_{_l}v \right)^2
}\frac{df_{\alpha }}{dp}\right) dp}{4\pi\sum\limits_{\alpha}\omega_{p\alpha
}^{2}\oint \frac{1}{\left( v-v_{g}\right) }\frac{df_{\alpha}}{dp}dp}\times
\frac{\partial ^{2}}{\partial \xi ^{2}}\mid E_{\parallel}^{(1)}\mid ^{2},
\label{Def_rho_1}
\end{equation}
where $E_{\parallel}^{(1)}$ is defined by the equation (\ref{solution}). Let us
note that this is a solution of the kinetic equation corresponding to the
slowest mode only ($l=0$ in eq.[\ref{sum}]), thus describing a charge
distribution within an envelope soliton. Since the distribution functions of
electrons $f_e$ and positrons $f_{p}$ appearing in equation (\ref{Def_rho_1})
are not equal, the resulting charge density contrast associated with the
soliton will be substantially  non-zero.

It is known that if $q\,G> 0$ \citep[the so-called Lighthill
condition;][]{lghill}, then the equation (\ref{solution}) describes a
soliton-like solution. For pulsar magnetospheric  plasma the coefficients of
equation (\ref{shro}) are
\begin{eqnarray}
G &=&\frac{1}{4}\frac{\gamma_{_p}^{2}c^{2}}{\omega_{p}}G_{d},  \nonumber \\
q &=&\left( \frac{e}{m_{e}c}\right) ^{2}\frac{1}{\gamma_{_p}^{2}\omega_{p}}
q_{_d},  \nonumber \\
s &=&\left( \frac{e}{m_{e}c}\right) ^{2}\frac{1}{\gamma_{_p}^{2}\omega_{p}}
s_{_d},
\label{G_s_q}
\end{eqnarray}
in the LFR, where $G_{d},q_{_d}$ and $s_{_d}$ depend on the distribution
function of plasma as shown in Figure \ref{fig3}. Obviously, the product $q\,G>
0$ for a wide range of parameters, so the Langmuir high frequency oscillations
modulated by the low-frequency beatings resulting from the range of linear wave
frequencies, will evolve into the solitary waves. We are modelling the
distribution function as
\begin{equation}
f_{\alpha }\thicksim \exp \left( -\left( \frac{p-p_{_\alpha }}{p_{_T}}
\right) ^{2}\right),
\label{distr_func}
\end{equation}
where $p_{_\alpha }$ for $p_{_e} \approx \gamma_{_p}$. In numerical
calculations we are using $p_{_p}/\gamma_{_p}=\left( 0.5 - 2\right)$,
$p_{_T}/\gamma_{_p}=\left( 0.5 - 1.5\right)$ and dimensionless momentum
$p_{_T}$ describes a degree of plasma thermalization. The maximum growth rate
of modulational instability can be written as
\begin{equation}
\Gamma_{m}\thicksim 2.7\times 10^{7}R_{_{50}}^{-1.5}~\left(
\frac{\dot{P}_{_{-15}}}{P}\right)^{0.25}
\gamma_{_2}^{-1.5}~\left( q_{_d}^{2}+s_{_d}^{2}\right)^{\frac{1}{2}}~
\chi ,
\label{Non_l_gr}
\end{equation}
where
\begin{equation}
\chi \equiv \frac{\left| E_{\parallel o}^{(1)}\right| ^{2}}{4\pi \gamma
_{p}m_{e}c^{2}n_{o}}\ll 1
\label{Dim_amp}
\end{equation}
which describes a ratio of energy densities of Langmuir waves and plasma. The
nonlinear growth rate becomes equal to the linear one when $\chi \thicksim
0.04.$ Typically $\chi\thicksim 0.1$, so the instability can easily develop
with growth rate high enough to satisfy equation (\ref{Cl}). The linear
instability is of the hydrodynamical type, meaning that half of the particle's
energy can be transferred to the plasma turbulence.

\section{Coherent curvature radiation}

In order to calculate a power of the coherent curvature radiation of the
soliton modeled as a  system of three coupled charged bunches presented in
Figure \ref{fig3}, let us express the current in the form
\begin{equation}
\mathbf{J}\left( \mathbf{r},t\right) =Q\left( \frac{d\mathbf{r}_{0}}{dt}
\delta \left( \mathbf{r}-\mathbf{r}_{0}\right) -\frac{1}{2}\frac{d\mathbf{r}
_{1}}{dt}\delta \left( \mathbf{r}-\mathbf{r}_{1}\right) -\frac{1}{2}\frac{d
\mathbf{r}_{2}}{dt}\delta \left( \mathbf{r}-\mathbf{r}_{2}\right) \right),
\label{Jrt}
\end{equation}
where $\mathbf{r}_{0}$ is the radius vector of the central bunch, and
$\mathbf{r}_{1}$ and $\mathbf{r}_{2}$ are the radius vectors of the side
bunches, respectively. The Cartesian components of radius vectors are
\begin{eqnarray}
r_{0x}=r_{c}\cos\left( \omega_{o}t\right), \qquad \qquad
r_{0y}&=&r_{c}\sin\left(\omega_{o}t\right),  \nonumber \\
r_{1x}=r_{c}\cos\left( \omega_{o}t+\varphi_{o}\right), \qquad \qquad
r_{1y}&=&r_{c}\sin\left( \omega_{o}t+\varphi_{o}\right),  \nonumber \\
r_{2x}=r_{c}\cos\left( \omega_{o}t-\varphi_{o}\right),\qquad \qquad
r_{2y}&=&r_{c}\sin\left( \omega_{o}t-\varphi_{o}\right),
\label{Buncor}
\end{eqnarray}
where $r_{c}$ is the radius of an effective circular orbit, $\varphi_{o}=\Delta
/r_{c}$ is the angle between the central and peripheral particle's radius
vectors, and $\omega_{_0} = v/r_{c}$ is the particles angular velocity (an
effective gyro-frequency). The origin of the coordinate system is chosen in the
center of the curvature, $z$-axis is directed perpendicular to the plane of the
curvature, and $y$-axis is tangent and $x$-axis is perpendicular to the local
magnetic field, respectively. In our case $r_{c}=\mathcal{R}$ in the radiation
generation region, where the radius of curvature of dipolar field lines
\begin{equation}
\mathcal{R}\thicksim 7\times 10^{8}R_{_{50}}^{1.5} ~~{\rm cm}.
\end{equation}
According to a well-known method \citep[e.g.,][]{ll62} we can calculate
electromagnetic field of the charged system far from it, in the so-called `wave
zone'. Let us start with the standard determinations of the current Fourier
components and vector potential. The current is
\begin{equation}
\mathbf{J}\left( \mathbf{r},t\right) =\sum_{n=-\infty }^{\infty }\mathbf{J}
_{n}\left( \mathbf{r}\right) e^{-i\omega_{_0}nt},
\label{B1} \end{equation}
where
\begin{equation}
\mathbf{J}_{n}\left( \mathbf{r}\right) =\frac{1}{T}\int\limits_{-T/2}^{T/2}
\mathbf{J}\left( \mathbf{r},t\right) e^{i\omega_{_0}nt} dt.
\label{B2}
\end{equation}
The Fourier component of the vector-potential is defined as
\begin{equation}
\mathbf{A}_{n}=\frac{\exp \left( ikR_{0}\right) }{cR_{0}T}
\int\limits_{-\infty }^{\infty }\int\limits_{-T/2}^{T/2}\mathbf{J}\left(
\mathbf{r},t\right) e^{i(\omega_{_0}nt-\mathbf{kr})} d\mathbf{r}dt,
\label{An}
\end{equation}
where $R_{0}$ is a distance from the radiating system to observer. In our case
the current is just a motion of three small charged `bunches'. The bunches are
moving along the circular trajectory. In this configuration the central bunch
has a charge $Q$ and two other have the charge $-\frac{1}{2}Q$ each. This
configuration is symmetric with respect of the central bunch. The axes are
chosen as follows: The particles' trajectory lies in the plane $X0Y$, $z$-axis
is perpendicular of this plane. $\mathbf{k}$-vector lies in the plane $Z0Y$.
Consequently from equations (\ref{Jrt}) and (\ref{An}) we obtain
\begin{equation}
\mathbf{A}_{n}=Q\frac{\exp\left(ikR_{0}\right)}{cR_{0}T}
{}~\int\limits_{-T/2}^{T/2} \left( \frac{d\mathbf{r}_{0}}{dt}
e^{i\left(\omega_{_0}nt-\mathbf{kr}_{0}\right)} -\frac{1}{2}\frac{d\mathbf{r}
_{1}}{dt}e^{i\left(\omega_{_0}nt-\mathbf{kr}_{1}\right)} -\frac{1}{2}\frac{d
\mathbf{r}_{2}}{dt}e^{i\left(\omega_{_0}nt-\mathbf{kr}_{2}\right)} \right)dt,
\label{An2}
\end{equation}
where
\begin{equation}
\mathbf{kr}_{0}=kr_{c}\cos(\theta )\sin\left(\omega_{_0}t\right);
\;
\mathbf{kr}_{1}=kr_{c}\cos(\theta)\sin\left(\omega_{_0}t+\varphi_{o}\right);
\;
\mathbf{kr}_{2}=kr_{c}\cos(\theta )\sin\left(\omega_{_0}t-\varphi_{o}\right).
\label{k_r}
\end{equation}
Here $\theta$ is the angle between the $\mathbf{k}$-vector and $y$-axis,
$T=2\pi/\omega_{_0}$, and we assume that $v \thicksim c,$. Therefore, we have
from equation (\ref{An2}) for $x$ and $y$ components
\begin{eqnarray}
\mathbf{A}_{xn} &=&-Qr_{c}\omega_{_0}\frac{\exp \left( ikR_{0}\right)}{
cR_{0}T}\int\limits_{-T/2}^{T/2}\{\sin\left( \omega_{_0}t\right) \exp
\left[i\omega_{_0}nt-ikr_{c}\cos(\theta)\sin\left(\omega_{_0}t\right)
\right] -  \nonumber \\
&&-\frac{1}{2}\sin\left( \omega_{_0}t+\varphi_{o}\right) \exp\left[
i\omega_{_0}nt-ikr_{c}\cos(\theta)\sin\left(\omega_{_0}t+\varphi
_{o}\right)\right] -  \nonumber \\
&&-\frac{1}{2}\sin\left( \omega_{_0}t-\varphi_{o}\right) \exp\left[
i\omega_{_0}nt-ikr_{c}\cos(\theta )\sin\left(\omega_{_0}t-\varphi
_{o}\right)\right]\}dt.
\end{eqnarray}
\begin{eqnarray}
\mathbf{A}_{yn} &=&Qr_{c}\omega_{_0}\frac{\exp\left(ikR_{0}\right)}{
cR_{0}T}\int\limits_{-T/2}^{T/2}\{\cos\left(\omega_{_0}t\right) \exp\left[
i\omega_{_0}nt-ikr_{c}\cos(\theta )\sin\left( \omega_{_0}t\right)
\right] -  \nonumber \\
&&-\frac{1}{2}\cos\left( \omega_{_0}t+\varphi_{o}\right) \exp\left[
i\omega_{_0}nt-ikr_{c}\cos(\theta )\sin\left( \omega_{_0}t+\varphi
_{o}\right)\right] -  \nonumber \\
&&-\frac{1}{2}\cos\left( \omega_{_0}t-\varphi_{o}\right) \exp\left[
i\omega_{_0}nt-ikr_{c}\cos(\theta )\sin\left( \omega_{_0}t-\varphi
_{o}\right)\right] \}dt.
\end{eqnarray}
Introducing the following variables
\begin{equation}
\varphi =\omega_{_0}t,\hspace{0.5in}z=kr_{c}\cos(\theta )
\label{B11}
\end{equation}
and using
\begin{equation}
\int\limits_{-\pi }^{\pi }\left\{ \sin\left( \varphi \pm \varphi
_{o}\right) \exp \left[ in\varphi -iz\sin\left( \varphi \pm \varphi
_{o}\right) \right] \right\} d\varphi =2\pi i\exp \left( \mp in\varphi
_{o}\right) J_{n}^{\prime }\left( z\right) ;
\label{B12}
\end{equation}
\begin{equation}
\int\limits_{-\pi }^{\pi }\left\{ \cos\left( \varphi \pm \varphi_{o}\right)
\exp \left[ in\varphi -iz\sin\left( \varphi \pm \varphi_{o}\right) \right]
\right\} d\varphi =2\pi \exp \left( \mp in\varphi_{o}\right)
\frac{n}{z}J_{n}\left( z\right) ,
\label{B13}
\end{equation}
we obtain the components of the vector potential in the form
\begin{equation}
\mathbf{A}_{xn}=-2\pi i\left( Qr_{c}\frac{\exp \left( ikR_{0}\right) }{
cR_{0}T}\right) \left[ 1-\cos\left( n\varphi_{o}\right)\right]
J_{n}^{\prime }\left( z\right) ;
\label{B14}
\end{equation}
\begin{equation}
\mathbf{A}_{yn}=2\pi \left( Qr_{c}\frac{\exp \left( ikR_{0}\right) }{cR_{0}T}
\right) \left[1-\cos\left( n\varphi_{o}\right)\right] \frac{n}{z}J_{n}\left(
z\right),
\label{B15}
\end{equation}
where $J_{n}\left( z\right)$ is the Bessel function of the first order. For the
radiation intensity with frequency $\omega =n\omega_{_0}$ emitted within the
solid angle $d\sigma $ we have
\begin{equation}
dI_{n}=\frac{c}{2\pi }\left| \mathbf{k\times A}_{n}\right|^{2}R_{0}^{2}d\sigma
,
\label{B16}
\end{equation}
where
\begin{equation}
\left| \mathbf{k\times A}\right|
^{2}=A_{x}^{2}k^{2}+A_{y}^{2}k^{2}\sin^{2}\left( \theta \right) .
\label{B17}
\end{equation}
So finally we obtain
\begin{eqnarray}
dI_{n} &=&\frac{c}{2\pi }\left[ A_{xn}^{2}k^{2}+A_{yn}^{2}k^{2}\sin^{2}\left(
\theta \right) \right] R_{0}^{2}d\sigma =  \nonumber \\
&=&2\pi ck^{2}\left( \frac{Qr_{c}}{cT}\right) ^{2}\left[1-\cos\left(
n\varphi_{o}\right)\right]^{2}\left( J_{n}^{\prime }\left( z\right) ^{2}+
\frac{n^{2}}{z^{2}}J_{n}\left( z\right) ^{2}\sin^{2}\left( \theta \right)
\right) d\sigma.
\label{B18}
\end{eqnarray}
Taking into account that $\omega_{_0}=v/r_{c}$, $T=2\pi /\omega_{_0}$ and
$kc=n\omega_{_0}$, we find
\begin{equation}
dI_{n}=\frac{n^{2}c}{2\pi r_{c}^{2}}Q^{2}\left[1-\cos\left(
n\varphi_{o}\right)\right]^{2}\left\{ J_{n}^{\prime }\left( z\right)
^{2}+\left(
\frac{n}{z}\right)^{2}J_{n}\left( z\right) ^{2}\sin^{2}\left( \theta
\right) \right\} d\sigma.
\label{B19}
\end{equation}
Using $v \thicksim c$ and writing
\begin{equation}
\frac{n}{z}=\frac{n}{kr_{c}\cos(\theta )}\thickapprox \frac{1}{\cos(\theta)},
\label{B20}
\end{equation}
where we make the approximation $z=kr_{c}\cos(\theta )\thickapprox n
\cos(\theta )$, we obtain
\begin{equation}
dI_{n}=Q^{2}\frac{n^{2}\omega_{_0}^{2}}{2\pi c}\left[ 1-\cos\left(
n\varphi_{o}\right) \right]^{2}\left\{ J_{n}^{\prime }\left( z\right)
^{2}+\tan^{2}\left( \theta \right) J_{n}\left( z\right) ^{2}\right\} d\sigma .
\label{dInoc}
\end{equation}
Since the part of the above equation in the square brackets does not depend on
the solid angle $\sigma$ and the integral of the part in the braces is well
known in the relativistic case \citep{ll62}, the integral of (\ref{dInoc}) is
straightforward. Using $\omega = kc$, $I_{\omega } = I_{n}dn = I_{n} d\omega
/\omega_{_0}$, and
\begin{equation}
n\varphi_{o}=\frac{\omega}{\omega_{_0}}\frac{\Delta}{r_c}=\frac{3}{2}\gamma_{_o}^3
\frac{\Delta}{{\cal R}}\frac{\omega}{\omega_c}=a\frac{\omega}{\omega_c},
\label{B22}
\end{equation}
where $a=\gamma_{_0}^3\Delta/{\cal R}$ and $\varphi_0$ is marked in Figure
\ref{fig3}, we can integrate expression (\ref{dInoc}) which leads to equation
(\ref{Eq12}) in the main body of the paper describing a spectral power emitted
by solitons. From the condition for coherency $\Delta\ll\lambda/2$, where
$\lambda\sim\pi{\cal R}/\gamma_{_0}^3$ is a wavelength of the coherent
radiation emitted by a soliton with characteristic size $\Delta$, it follows
that $a\ll\pi/2$.

\vskip
12pt~\scalebox{0.5}{\includegraphics*{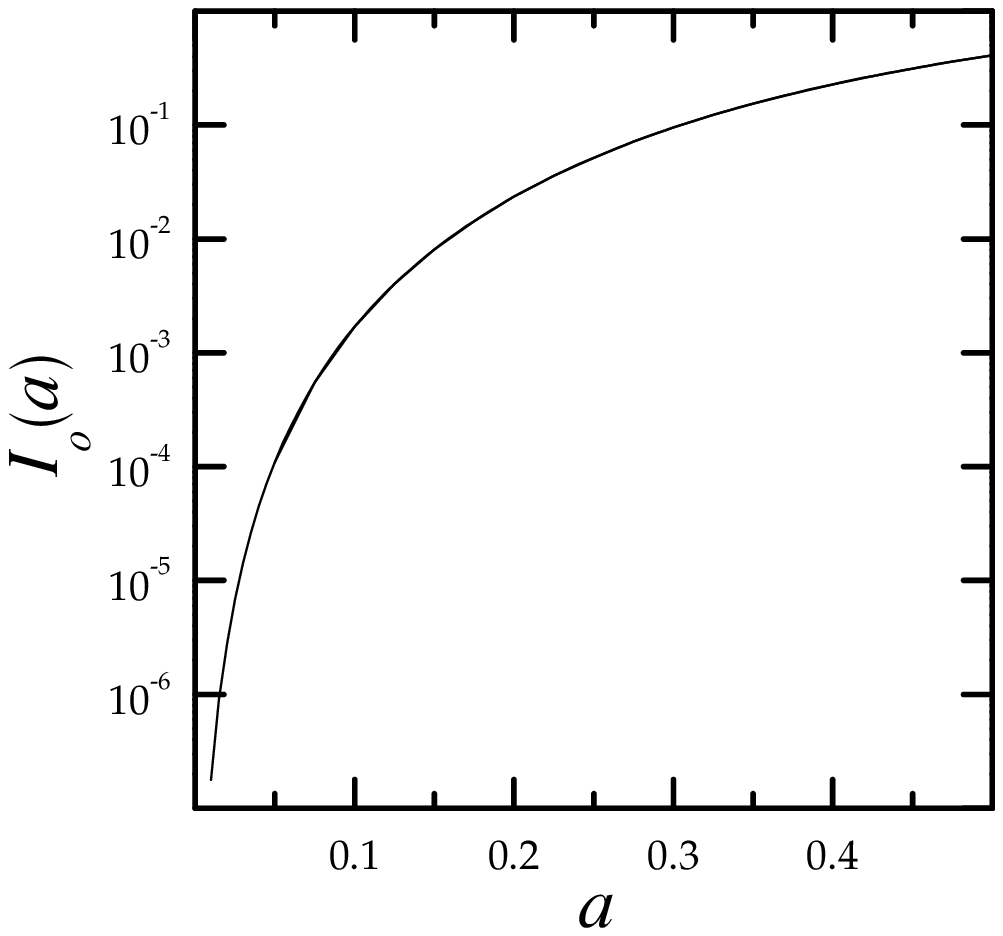}}~\figcaption[MGPFIG5.EPS]{Values of the integral $I_{o}(a)$ (eq.[\ref{L1}]) versus parameter $a$ (eq.[\ref{a}]). \label{fig5}}~\vskip 2pt

\section{Definitions\label{Defs}}
\noindent
$a=\gamma_{_0}^3\Delta/{\cal R}$ - dimensionless parameter (see eq.[\ref{a}]).
\\
$B_{12}=B_0/10^{12}$ G - value of the surface magnetic field in units of
$10^{12}$ Gauss. \\
$D$ - characteristic perpendicular spark dimension in cm (approximately equal
to $h$).\\
$\Delta$ - characteristic soliton size along dipolar field lines (see
eq.[\ref{Delta}]). \\
$\Delta_d$ - dimensionless parameter shown in Figure \ref{fig1}. \\
$\Delta \gamma $ - difference between the average Lorentz factors of electrons
and positrons. \\
$\delta_{\omega}\thicksim0.5$ - dimensionless parameter calculated in
\citet{am98}. \\
$f_{\alpha}$ - distribution function of type $\alpha$ plasma particles (see
eq.[\ref{distr_func}]). \\
$G$ - cefficient of equation \ref{shro} (see eq.[\ref{Def_G}]).\\
$G_{d}$ - dimensionless parameter shown in Figure \ref{fig1}
(eq.[\ref{G_s_q}]).\\
$\Gamma_l$ - linear growth rate (see eq.[\ref{Gl}]). \\
$\Gamma_{m}$ - growth rate of modulational instability (see eq.[\ref{gamma}]).
\\
$\gamma_{_0} =(1-v_{g}^2/c^2)^{-1/2}$ - Lorentz factor corresponding to
solitons (see eq.[\ref{coord}]). \\
$\gamma_{_b}\sim 10^6$ - average Lorentz factor of the primary beam particles
(see eq.[\ref{Eq1}]). \\
$\gamma_{_p}\sim 100$  - average Lorentz factor of the plasma particles. \\
$\gamma_{_2}=\gamma_{_p}/100$. \\
$\gamma_{_T}\thicksim p_{_T}$ - characteristic thermal spread of the plasma
particles (eq.[\ref{distr_func}]). \\
$h\simeq 5\times 10^{3}~\mathcal{R}_{6}^{2/7}~B_{12}^{-4/7}~P^{3/7}$~cm  -
polar gap height. \\
$I_{\omega}$ - spectral power of coherent curvature radiation of a soliton (see
eq.[\ref{Iom}]). \\
$\varkappa\sim\gamma_{_b}/\gamma_{p}$ - Sturrock's multiplication factor (for
typical pulsar $\varkappa\thicksim10^4$). \\
$\varkappa_4=\varkappa/10^4$, $\varkappa_4\thicksim1.6/\gamma_{_2}$ (see
eq.[\ref{Eq1}]). \\
$L_1$ - power radiated by a single soliton (see eq.[\ref{L1}]). \\
$L_m$ - maximum kinetic spark luminosity (see eq.[\ref{Eq24}])\\
$L_{sd}$ - pulsar spin-down luminosity. \\
$L_t$ - total power radiated by all solitons (see eq.[\ref{Lt}]). \\
$N_{sl}$ - number of solitons associated with a single spark (see
eq.[\ref{Nsl}]). \\
$N_{sp}$ - number of sparks on the polar cap (see eq.[\ref{Nsp}]). \\
$N_t$ - total number of solitons (see eq.[\ref{Nt}]). \\
$\nu_m$ - characteristic (maximum) frequency of the soliton curvature radiation
(see eq.[\ref{nm}]). \\
$\omega_{o}=v/r_c$ - angular velocity of a particle. \\
$\omega_c=1.5~\omega_{o}\gamma_{_0}^3$ - characteristic frequency of single
particle curvature radiation. \\
$\omega_{_l}\simeq2\delta_{\omega}\gamma_{_p}\omega_{_p}$ - characteristic
frequency of the excited Langmuir waves (see eq.[\ref{wl}]). \\
$\omega_{_p}=\left(4\pi e^2n/m_e\right)^{1/2}$ - plasma frequency (see
eq.[\ref{wp}]).\\
$P$ - pulsar period in seconds. \\
$\dot{P}_{_{-15}}$ - period derivative in units of $10^{-15}~\rm{s~s^{-1}}$. \\
$Q$ - charge of the central bunch in the soliton wavelet (see Fig.\ref{fig3}).
\\
$Q_d$ - dimensionless parameter shown in Figure \ref{fig1}. \\
$q$ - coefficient of local-nonlinear term in equation \ref{shro} (see
eq.[\ref{Def_q}]). \\
$q_{_d}$ - dimensionless parameter shown in Figure \ref{fig1} (see
eq.[\ref{G_s_q}]). \\
$R_{_{50}}=r/(50R)$ - distance from the stellar surface in 50 stellar radii
$R=10^6$~cm. \\
$\mathcal{R}\sim 7\times 10^8R_{50}^{1.5}$~cm - curvature radius of the dipolar
magnetic field lines. \\
$\mathcal{R}_6={\cal R}/{R}$ - curvature radius of the magnetic field lines at
the polar cap region in units of $10^6$~cm. \\
$r$ - radial coordinate (absolute value of radius vector). \\
$r_c\sim{\cal R}$ - curvature radius of particles trajectory. \\
$r_{_{in}}\simeq R_{_{50}}\,R$ - linear instability altitude. \\
$r_{_p}\sim10^4P^{-0.5}~{\rm cm}$ - polar cap radius. \\
$\Delta r$ -  characteristic longitudinal dimension of linear instability
region. \\
$\rho$ - slowly-varying charge density inside soliton (see eq.[\ref{Eq8}]). \\
$\rho_{d}$ - dimensionless parameter shown in Figure \ref{fig1}. \\
$S_{\perp}$ - cross-section of spark-associated soliton (see eq.[\ref{Eq15}]).
\\
$s$ - coefficient of non-local nonlinear term in equation \ref{shro}
(eq.[\ref{Def_s}]). \\
$s_{_d}$ - dimensionless parameter shown in Figure \ref{fig1}
(eq.[\ref{G_s_q}]). \\
$V$ - canonical maximum potential drop within the gap in Volts. \\
$\mathcal{V}$ - volume of spark-associated soliton (see eq.[\ref{Eq16}]). \\
$v_{g}=\partial\omega_{_l}/\partial k_{_l}$ - group velocity of Langmuir waves
in LFR. \\
$y=\gamma_{_0}/\gamma_{_p}$ - dimensionless parameter shown in Figure
\ref{fig1}. \\

\clearpage

\begin{deluxetable}{llllllll}
\footnotesize
\tablecaption{Observed and calculated pulsar luminosities.\label{tbl1}}
\tablewidth{0pt}
\tablehead{
\colhead{PSR} & \colhead{$P$} [s] & \colhead{$\dot{P}_{-15}$} &
\colhead{$L_{sd}$} [erg/s] & \colhead{$L_{_R}$} erg/s & \colhead{$L_{t}$}
[erg/s] &  \colhead{$y$} & \colhead{$\gamma_{_2}$}}
\startdata
B0531$+$21 & $0.033$ & $421$ & $4.6\times 10^{38}$ & $1.3\times 10^{29}$ &
$1.7\times 10^{29}$ &  $1.5$ & $1$ \\
B0833$-$45 & $0.0893$ & $125$ & $6.9\times 10^{36}$ & $4.3\times 10^{28}$
&$8.0\times 10^{28}$ & $1.6$ & $1$ \\
B1610$-$50 & $0.232$ & $493$ & $1.6\times 10^{36}$ & $2.8\times 10^{28}$ &
$7.7\times 10^{28}$ &  $2$ & $1$ \\
B0950$+$08 & $0.253$ & $0.229$ & $5.6\times 10^{32}$ & $2.3\times 10^{26}$&
$6.6\times 10^{26}$ & $1.5$ & $0.65 $ \\
B1133$+$16 & $1.188$ & $3.73$ & $8.8\times 10^{31}$ & $7.4\times 10^{26}$
&$9.6\times 10^{26}$ & $2$ & $0.9$ \\
B1746$-$30 & $0.61$ & $7.9$ & $1.4\times 10^{33}$ & $3.0\times 10^{28}$ &
$9.4\times 10^{28}$ & $2.2$ & $1$ \\
B0525$+$21 & $3.75$ & $40$ & $3.0\times 10^{31}$ & $1.1\times 10^{28}$ &
$1.5\times 10^{28}$ & $2$ & $1.2$ \\
J2144$-$39$^*$ & $8.51$ & $0.48$ & $3.0\times 10^{28}$ & $5.0\times 10^{24}$ &
$7.2\times 10^{24}$ & $2.2$ &  $0.9$\\
\enddata
\tablecomments{PSR name, $P$ - pulsar period in seconds, $\dot{P}_{-15}$ -
period derivative in $10^{-15}\rm{s~s^{-1}}$, $L_{sd}=3.8\times
10^{31}\dot{P}_{-15}P^{-3}$~erg~s$^{-1}$ - spindown luminosity,
$L_{R}=3.5\times 10^{25+x}$~erg~s$^{-1}$ - observed radio luminosity, where
$x={\mathrm Log}\,L$ in mJy~kpc$^2$ from the pulsar catalog, $L_t$ - total
pulsar luminosity calculated from equation (\ref{Lt}), parameters $y$ and
$\gamma_{_2}$ used in calculations (see Appendix C and Figure 1 for
explanations). $*$ - data for PSR J$2144-3933$ are taken from \citet{ymj99}}.
\end{deluxetable}


\begin{thebibliography}{}
\bibitem[Asseo(1993)]{ass93}  Asseo, E. 1993, \mnras, 264, 940
\bibitem[Asseo \& Melikidze(1998)]{am98}  Asseo, E., \& Melikidze, G. I. 1998,
\mnras, 301, 59
\bibitem[Blaskiewicz et al.(1991)]{bcw91}  Blaskiewicz, M., Cordes, J. M., \&
Waserman, I. 1991, \apj, 370, 643
\bibitem[Cheng et al.(1998)]{cgz98} Cheng, K. S., Gil, J., Zhang, L. 1998,
\apj, 493, L35
\bibitem[Cheng \& Zhang(1999)]{cz99} Cheng, K. S., Zhang, L. 1999, \apj, 515,
337
\bibitem[Cheng \& Ruderman(1977a)]{cr77a}  Cheng, A. F., \& Ruderman, M. A.
1977a, \apj, 212, 800
\bibitem[Cheng \& Ruderman(1977b)]{cr77b}  Cheng, A. F. \& Ruderman, M. A.
1977b, \apj, 214, 598
\bibitem[Cheng \& Ruderman(1980)]{cr80}  Cheng, A. F. \& Ruderman, M. A. 1980,
\apj, 235, 576
\bibitem[Cordes(1978)]{cor78}  Cordes, J. M. 1978, \apj, 222, 1006
\bibitem[Cordes (1992)]{cor92}  Cordes, J. M. 1992, in IAU Colloq. 128, The
Magnetospheric Structure and Emission Mechanism of Radio Pulsars, ed. T. H.
Hankins, et al. (Zielona G\'ora, Poland: Pedagogical Univ. Press), 253
\bibitem[Deshpande \& Rankin (1999)]{dr99}  Deshpande, A. A., \& Rankin, J. M.
1999, \apj, 524, 1008
\bibitem[Gangadhara(1997)]{gan97}  Gangadhara, R. T. 1997, \aap, 327, 155
\bibitem[Gangadhara(1999)]{gan99}  Gangadhara, R. T. 1999, \aap, 392, 474
\bibitem[Gaponov \& Miller(1958)]{gm58}  Gaponov, A. V., \& Miller, M. A. 1958,
Soviet Phys. JETP, 34, 242
\bibitem[Gil et al.(1995)]{gkms95}  Gil, J., Kijak, J., Maron, O., \& Sendyk,
M. 1995, \aap, 301, 177
\bibitem[Gil et al.(1993)]{gkz93}  Gil, J., Kijak, J., \& Zycki, P. 1993, \aap,
272, 207
\bibitem[Gil \& Lyne(1995)]{gl95} Gil, J., \& Lyne, A.G., 1995, \mnras, 276,
L55
\bibitem[Gil et al.(1997)]{gkm97}  Gil, J., Krawczyk, A., \& Melikidze, G. I.
1997, Banach Center Publications, vol.41/2,  (Warsaw: Polish Academy of
Sciences), 239
\bibitem[Gil \& Mitra(1999)]{gm99} Gil, J., \& Mitra, D. 1999, \aap, submitted
\bibitem[(Paper I)]{gs00} Gil, J., \& Sendyk, M. 2000, \apj (Paper I)
\bibitem[Gil \& Snakowski(1990a)]{gsn90a}  Gil, J., \& Snakowski, J. K. 1990a,
\aap, 234, 237
\bibitem[Gil \& Snakowski(1990b)]{gsn90b}  Gil, J., \& Snakowski, J. K. 1990b,
\aap, 234, 269
\bibitem[Goldreich \& Julian(1969)]{gj69}  Goldreich, P., \& Julian, H. 1969,
\apj, 157, 869
\bibitem[Ichikawa \& Taniuti(1973)]{it73}  Ichikawa, Y. H., \& Taniuti, T. 1973
J. Phys. Soc. Japan, 34, 513
\bibitem[Ichikawa et al.(1973)]{ist73}  Ichikawa, Y. H., Suzuki, T., \&
Taniuti, T. 1973, J. Phys. Soc. Japan, 34, 1089
\bibitem[Karpman et al.(1975)]{karp75}  Karpman, V. I., Norman, C.A., ter Haar,
D., \& Tsitovich, V. N. 1975, Phys. Scripta, 11, 271
\bibitem[Karpman \& Krushkal(1969)]{kk69}  Karpman, V. I., \& Krushkal, E. M.
1969, JETP, 28, 277
\bibitem[Kazbegi et al.(1991)]{kmm91} Kazbegi, A.Z., Machabeli, G.Z., \&
Melikidze, G.I. 1991, \mnras, 253, 377
\bibitem[Kazbegi et al.(1992)]{kmm92}  Kazbegi, A. Z., Machabeli, G. Z., \&
Melikidze, G. I. 1992, in IAU Colloq. 128, The Magnetospheric Structure and
Emission Mechanism of Radio Pulsars, ed. T. H. Hankins, et al. (Zielona G\'ora,
Poland: Pedagogical Univ. Press), 232
\bibitem[Kijak \& Gil(1997)]{kg97}  Kijak, J., \& Gil, J. 1997, \mnras, 288,
631
\bibitem[Kijak \& Gil(1998)]{kg98}  Kijak, J., \& Gil, J. 1998, \mnras, 299,
855
\bibitem[Landau \& Lifshitz(1962)]{ll62}  Landau, L. D., \& Lifshitz, E. M.
1962, Classical Theory of Fields (Oxford: Pergamon Press)
\bibitem[Lighthill(1967)]{lghill} Lighthill, M. J. 1967, Proc. Roy. Soc. A229,
28
\bibitem[Lominadze et al.(1986)]{lmmp86}  Lominadze, J. G., Machabeli, G. Z.,
Melikidze, G. I., \& Pataraya, A. D. 1986, Sov. J. Plasma Phys., 12, 712
\bibitem[Lutikov et al.(1999)]{lbm99}  Lutikov, M., Blandford, R. D., \&
Machabeli, G. 1999, \mnras, 305, 338
\bibitem[Lyne \& Manchester(1988)]{lm88}  Lyne, A. G., \& Manchester, R. N.
1988, \mnras, 234, 477
\bibitem[Machabeli(1991)]{mach91}  Machabeli, G. Z. 1991, Plasma Phys. and
Controlled Fusion, 33, 1227
\bibitem[Malofeev et al.(1994)]{mal94} Malofeev, V. M., Gil, J., Jessner A.,
Malov, I., Sieber, W., \& Wielebinski, R. 1994, \aap, 285, 201
\bibitem[Melikidze \& Pataraya(1980a)]{mp80a}  Melikidze, G. I., \& Pataraya,
A. D. 1980, \apss, 68, 49
\bibitem[Melikidze \& Pataraya(1980b)]{mp80b}  Melikidze, G. I., \& Pataraya,
A. D. 1980, Astrofizika, 16, 161
\bibitem[Melikidze \& Pataraya(1984)]{mp84}  Melikidze, G. I., \& Pataraya, A.
D. 1984, Astrofizika, 20, 157
\bibitem[Melrose \& Gedalin(1999)]{mg99} Melrose, D. B., \& Gedalin, M. E.
1999, \apj, 521, 351
\bibitem[Michel(1987)]{mich87}  Michel, F. C. 1987, \apj, 322, 822
\bibitem[Petviashvili(1976)]{petv76} Petviashvili, V.I. 1976, Sov. J. Plasma
Phys., 2, 247
\bibitem[Radhakrishnan \& Cooke(1969)]{rc69} Radhakrishnan V., \& Cooke D. J.
1969, Astrophys. Lett., 3, 225
\bibitem[Rankin(1983)]{r83} Rankin, J. M. 1983, \apj, 274, 33
\bibitem[Rankin(1990)]{r90} Rankin, J. M. 1990, \apj, 352, 314
\bibitem[Rankin(1993)]{r93} Rankin, J. M. 1993, \apj, 405, 285
\bibitem[Ruderman \& Sutherland(1975)]{rs75} Ruderman, M. A., \& Sutherland, P.
G. 1975, \apj, 196, 51
\bibitem[Sagdeev(1979)]{sag79} Sagdeev, R. Z. 1979, Rev. Mod. Phys., 51, 1
\bibitem[Sturrock(1971)]{stu71} Sturrock, P. A. 1971, \apj, 164, 529
\bibitem[Taylor et al.(1993)]{tml93} Taylor, J. H., Manchester, R. N., \& Lyne,
A. G. 1993, \apjs, 88, 529 (Pulsar Catalog)
\bibitem[Taniuti \& Yajima(1969)]{ty69}  Taniuti, T., \& Yajima, N. 1969, J.
Math. Phys., 10,1369
\bibitem[Ursov \& Usov(1988)]{usov2}  Ursov, V. N., \& Usov, V. V. 1988, \apss,
140, 325
\bibitem[Usov(1987)]{usov1}  Usov, V. V. 1987, \apj, 320, 333
\bibitem[Weatherall(1997)]{w97} Wheatherall, J.C. 1997, \apj, 483, 402
\bibitem[Weatherall(1998)]{w98} Wheatherall, J.C. 1998, \apj, 506, 341
\bibitem[Xu et. al(1999)]{xqz99} Xu, R.X., Qiao, G.J., Zhang, B. 1999, \apj,
522, L109
\bibitem[Young et al.(1999)]{ymj99} Young, M.D., Manchester, R.N., \& Johnston,
S. 1999, Nature, 400, 848
\bibitem[Zakharov \& Shabat(1972)]{zs72} Zakharov, V. A., \& Shabat, A. B.
1972, Soviet Phys. JETP, 34, 62
\end{thebibliography}
\end{document}